\documentclass[12pt, a4paper]{article}
\usepackage{amsmath}
\usepackage{amsfonts}
\usepackage{amssymb}
\usepackage{graphicx}
\usepackage{mathptmx}
\usepackage{graphicx}
\usepackage{anysize}
\usepackage[font={scriptsize, sf, bf}, textfont=md, margin=1cm]{caption}
\date{\vspace{-5ex}}
\usepackage{sectsty}
\usepackage{ifpdf}
\usepackage{setspace}
\usepackage{fancyhdr}

\usepackage{lipsum} 
\sectionfont{\large}
\begin{document}
\title{\textbf{An Adaptive Threshold in Mammalian Neocortical Evolution}}
\author{Eric Lewitus$^1$$^{\infty}$, Iva Kelava$^1$$^{\infty}$, Alex T. Kalinka$^1$, \\Pavel Tomancak$^1$, and Wieland B Huttner$^1$$^*$}
\maketitle
{\scriptsize{$^1$Max Planck Institute of Molecular Cell Biology and Genetics, Pfotenhauerstr. 108, 01307 Dresden, Germany}}\\
\hspace*{0.5cm}{\scriptsize{$^{\infty}$These authors contributed equally\\\hspace*{0.5cm}$^*$Corresponding author}}\\
\setstretch{1.4}
\begin{abstract}
Expansion of the neocortex is a hallmark of human evolution. However, it remains an open question what adaptive mechanisms facilitated its expansion. Here we show, using gyrencephaly index (GI) and other physiological and life-history data for 102 mammalian species, that gyrencephaly is an ancestral mammalian trait. We provide evidence that the evolution of a highly folded neocortex, as observed in humans, requires the traversal of a threshold of $\sim$10$^9$ neurons, and that species above and below the threshold exhibit a bimodal distribution of physiological and life-history traits, establishing two phenotypic groups. We identify, using discrete mathematical models, proliferative divisions of progenitors in the basal compartment of the developing neocortex as evolutionarily necessary and sufficient for generating a fourteen-fold increase in daily prenatal neuron production and thus traversal of the neuronal threshold. We demonstrate that length of neurogenic period, rather than any novel progenitor-type, 
is sufficient to distinguish cortical neuron number between species within the same phenotypic group.
\end{abstract}
\pagebreak
\marginsize{2cm}{2cm}{1cm}{1cm}
\section*{Introduction}
Development of the human neocortex involves a lineage of neural stem and progenitor cells that forms a proliferative region along the ventricular epithelium. The proliferation of cells within this region expands the neocortex by increasing neuron number. At the onset of mammalian cortical neurogenesis, neuroepithelial cells transform into radially oriented apical radial glia (aRG), which proliferate extensively at the apical surface of the ventricular zone and divide asymmetrically to self-renew and generate a neuron, intermediate progenitor (IP), or basal radial glia (bRG) \cite{franco_shaping_2013}. IP cells delaminate from the apical surface and translocate their nucleus to the basal region of the ventricular zone (VZ) to form a second germinal layer, the subventricular zone (SVZ), where they divide symmetrically to generate two neurons \cite{noctor_neurons_2001,miyata_asymmetric_2004,haubensak_neurons_2004}. Similarly to aRG cells at the ventricular surface, bRG cells, which maintain a single fiber 
ascending only to the basal surface, divide asymmetrically \cite{fietz_osvz_2010,hansen_neurogenic_2010,shitamukai_oblique_2011,wang_new_2011}; but contrary to aRG cells, bRG in the human may both divide symmetrically and generate neurons via transit-amplifying progenitors (TAPs), a cell-type that is not observed to originate basally in the mouse \cite{hansen_neurogenic_2010}. The abventricular expansion of progenitors during cortical neurogenesis in humans further compartmentalizes the basal region into an inner and outer SVZ, driving the radial fibers to have divergent, rather than parallel, trajectories to the cortical plate, and thus creating the folded cortical pattern observed in gyrencephalic species through the tangential expansion of migrating neurons \cite{smart_unique_2002,borrell_emerging_2012}. For this reason, and based on supporting evidence obtained in the gyrencephalic human and ferret and lissencephalic mouse, it was originally thought that an abundance of asymmetrically dividing bRG 
cells in the outer SVZ was an evolutionary determinant for establishing a relatively large and gyrencephalic neocortex \cite{fietz_osvz_2010,hansen_neurogenic_2010,reillo_role_2011}. But recent work in the lissencephalic marmoset (\textit{Callithrix jacchus}) has shown that bRG cells may, in fact, exist in comparable abundance in both gyrencephalic and lissencephalic species \cite{garcia-moreno_compartmentalization_2012,kelava_abundant_2012} and so cannot alone be sufficient for either establishing or increasing cortical gyrification. Thus, despite considerable progress in the study of brain size evolution \cite{finlay_linked_1995,krubitzer_evolution_2005,hager_genetic_2012}, the adaptive mechanism that has evolved along certain mammalian lineages to produce a large and folded neocortex is not known. 

In this study, we analyzed physiological and life-history data from 102 mammalian species (Table S1; Table S2; External Database 1). We show that a gyrencephalic neocortex is ancestral to all mammals (Figure 1) and that GI (Figure S1), like brain size, has increased and decreased along many mammalian lineages. These changes may be reliably characterized by convergent adaptations into two distinct physiological and life-history programs (Figure 2a), resulting in a bimodal distribution of mammalian species (Figure 2b) with a robust threshold value for both GI and neuron number (Figure 3). Traversal of the threshold requires greater neuron production per gestation day (Figure 4a,b), which we argue is necessitated by the evolution of increased proliferative potential in  SVZ progenitors during cortical neurogenesis (Figure 5).

\begin{figure}
\centering
\makebox[\textwidth][c]{
\includegraphics[scale=0.70]{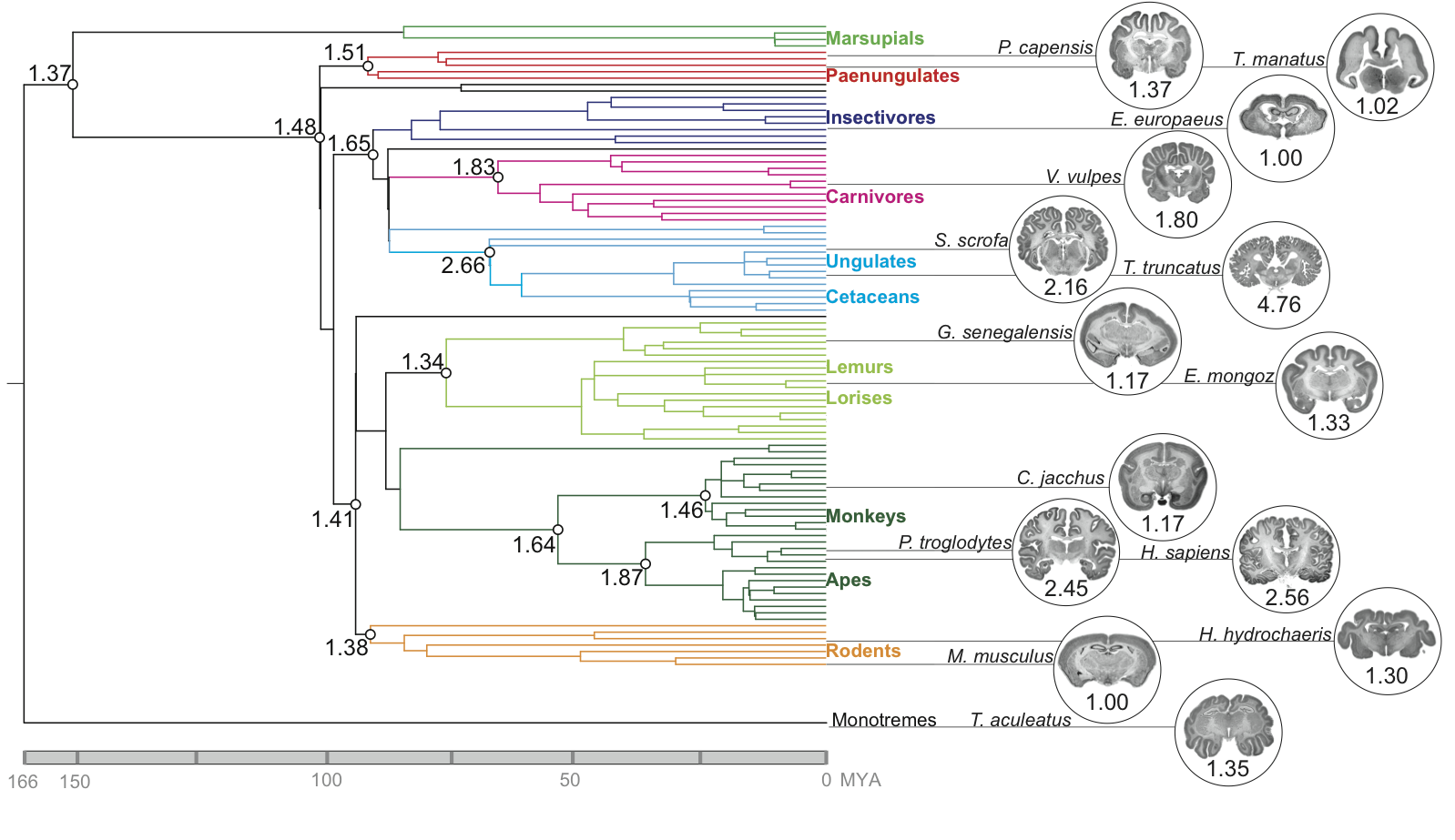} }
\caption{Ancestral reconstruction of GI values for 102 mammalian species. GI values were determined as illustrated in Figure S1 for the species listed in Table S1. Reconstructed GI values for putative ancestors are presented at selected internal nodes of the phylogenetic tree. MYA, million years ago; colors indicate taxonomic groups. Images of Nissl-stained coronal sections of representative species for each taxonomic group, downloaded from http://brainmuseum.org, along with respective GI values, are shown on the right.}
\label{Figure 1}
\end{figure} 

\section*{The mammalian ancestor was gyrencephalic}
We tested multiple evolutionary models for GI evolution.  The model that conferred most power to explain the GI values across the phylogeny while making the fewest assumptions about the data (i.e., had the lowest Akaike Information Criterion (AIC)) showed a disproportionate amount of evolutionary change to have occurred recently, rather than ancestrally, in mammals (Figure S2) and diverged significantly from a null model of stochastic evolution \cite{pagel_inferring_1999}. We identified a folded neocortex (GI =1.36 $\pm$ 0.16 s.e.m.) as an ancestral mammalian trait (Figure 1). It is apparent from ancestral and other internal node reconstructions (Figure S3) that GI is very variable, but also that reductions in the rate at which GI evolves have favored branches leading to decreases in GI (e.g., strepsirrhines and insectivores) and accelerations in that rate have favored branches leading to increases in GI (e.g., carnivores and caviomorphs). A simulation of the average number of total evolutionary transitions 
between GI values evidences more affinity for transitioning from high-to-low than low-to-high GI values: the majority of high-to-low transitions (58.3$\%$) occurred in species with a GI $<$ 1.47; and the fewest transitions (16.7$\%$) occurred across a threshold value of 1.5 (Figure S4). This indicates that, although there is an evident trend in mammalian history to become increasingly gyrencephalic, the most variability in GI evolution has been concentrated among species below a certain threshold value (GI = 1.5). We therefore present a picture of early mammalian history, contrary to those previously painted, but which is gathering evidence through novel approaches \cite{oleary_placental_2013,romiguier_genomic_2013}, that the Jurassic-era mammalian ancestor may, indeed, have been a large-brained species with a folded neocortex. 

\begin{figure}
\centering
\makebox[\textwidth][c]{
\includegraphics[scale=1.2]{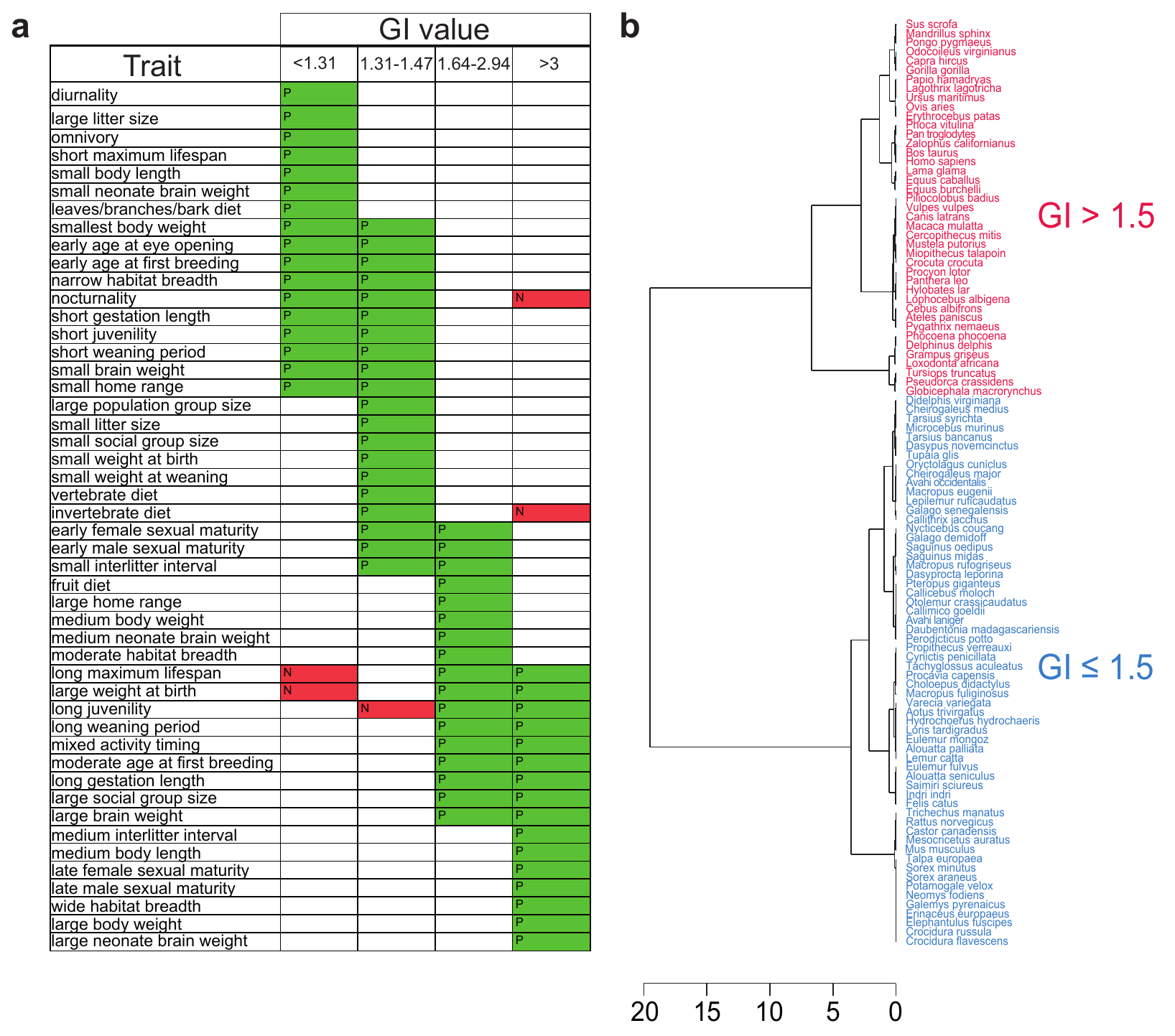}}
\caption{Clustering of GI values based on life-history association analysis (a) and minimum-energy distance (b). (a) Stochastic mapping of physiological and life-history traits with GI values for the 102 mammalian species listed in Table S1. GI values were separated into four groups based on clustering. Forty traits, each comprising 3 -– 6 character states, were analyzed (see Table S2 for a complete list), and the states showing a significant positive (P, green) or negative (N, red) association with a group of GI values are shown. Note the major overlap between the two low-GI groups (10/27) and between the two high-GI groups (9/24), whereas only 3/48 character states are shared between GI groups $\leq$ 1.5 versus $>$ 1.5.
(b) Hierarchical clustering based on minimum-energy distance of the GI values for 101 mammalian species. Note that the greatest clustering height is between species with GI values of $\leq$ 1.5 and $>$ 1.5.}
\label{Figure 2}
\end{figure}

\section*{A threshold in cortical neuron number}
The evolutionary effects of a folded neocortex on the behavior and biology of a species is not immediately clear. We therefore analyzed associations, across the phylogeny, of GI with discrete character states of 37 physiological and life-history traits (Table S2). Distinct sets of small but significant (R$^2$ $\leq$ 0.23, P $<$ 0.03) associations were found for species above and below a GI value of 1.5, indicating that these two groups of species adapt to their environments differently (Figure 2a). Both groups were sampled from across the phylogeny, showing no phylogenetic signal. Clustering analyses also supported a bimodal distribution above and below a threshold value of 1.5 (Figure 2b; Figure S5). To test the bimodal distribution explicitly, we regressed GI values against neuroanatomical traits and found that each scaling relationship could be explained comparably well by either a non-linear function (Figure 3a) or two grade-shifted linear functions, with the best-fit linear models drawing significantly 
different slopes (P = 3.4 x 10$^{-4}$) for high-GI ( $>$ 1.5) and low-GI ( $<$ 1.5) species. (Figure 3b,c). By plotting GI as a function of cortical neuron number, we were able to demarcate, with two significantly different linear regressions for high- and low-GI species (T = 4.611, d.f. = 29, P = 2.8 x 10$^{-4}$), a cortical no-man's-land centered on an area approximating 1 $\pm$ 0.11 x 10$^9$ neurons and 1.56 $\pm$ 0.06 GI (Figure 3d). The deviation of these results from previous work, which have shown strong phylogenetic signals associated with both GI \cite{pillay_order-specific_2007,zilles_development_2013} and neuron counts \cite{azevedo_equal_2009}, may be explained by our more than 2-fold increase in sampled species. Variation in GI, therefore, has not evolved linearly across the phylogeny, but has in fact been differentially evolved in two phenotypic groups. Each group may be characterized not only by a high ($>$ 1.5) or low ($<$ 1.5) GI value, but also by a distinct constellation of other 
physiological and life-history traits which have accompanied each group over evolutionary time.

\begin{figure}
\centering
\makebox[\textwidth][c]{
\includegraphics[scale=.7]{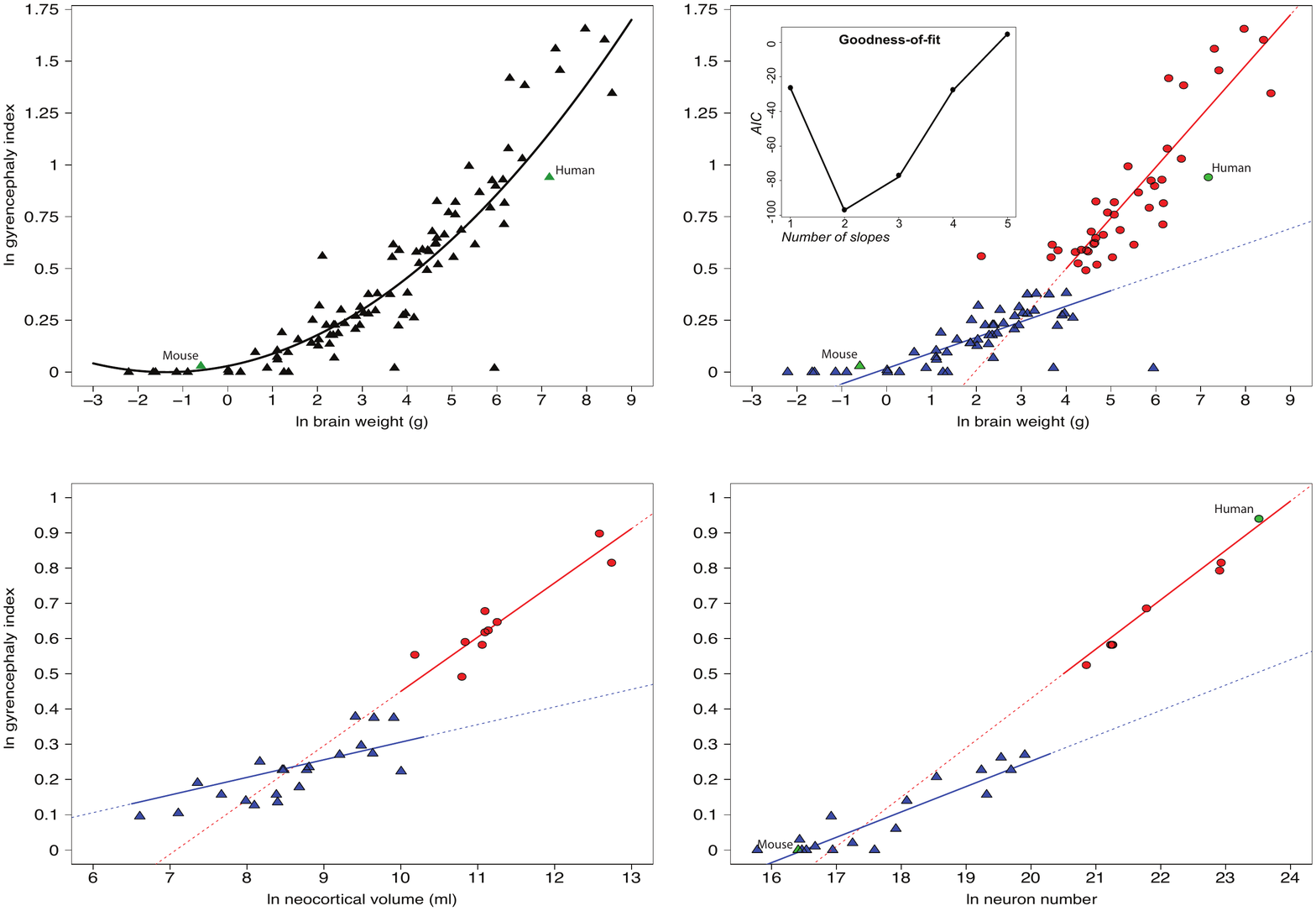}}
\caption{Ln-transformed plots showing GI values as a function of brain weight (a, b, 101 species), neocortical volume (c, 29 species) and cortical neuron number (d, 22 species). (a) Regression analysis using one non-linear fit for all values (y = 0.018x$^2$ + 0.037x + 0.014, R$^2$ = 0.612, P = 6 x 10$^{-5}$); (b-d) regression analyses using two different linear functions (b, blue line: y = 0.075x -– 0.481, R$^2$ = 0.56, P = 4 x 10$^{-5}$, red line: y = 0.245x + 0.018, R$^2$ = 0.73, P = 1 x 10$^{-5}$; c, blue line: y = 0.050x –- 0.194, R$^2$ = 0.21, P = 0.017, red line: y = 0.154x –- 1.09, R$^2$= 0.82, P = 0.004; d, blue line: y = 0.072x –- 1.188, R$^2$ = 0.81, P = 1 x 10$^{-4}$; red line: y = 0.140x -– 2.370, R$^2$ = 0.98, P = 3 x 10$^{-5}$) for species with GI values of $<$ 1.5 (blue triangles) and $>$ 1.5 (red circles), respectively; mouse and human are indicated by green symbols. The inset in (b) shows the AIC values for models fitted with 1 -- 5 linear slopes; note that a two-slope model best explains 
the data. See Table S1 for data.}
\label{Figure 3}
\end{figure} 

\section*{More efficient neurogenesis in large-brained species}
By establishing an evolutionary threshold based on both degree of gyrencephaly and neuron number, we identified two neurogenic phenotypic groups, which found support in their distinct life-history associations (see previous section). These groups could be further divorced by accounting for the amount of brain weight accumulated per gestation day -– confident proxies for neonate brain weight and neurogenic period, respectively (Figure S6) -– which we show to be, on average, 14-times greater in high- compared to low-GI species (Figure 4). Notably, each GI group is constituted by both altricial and precocial species, so the degree of pre- versus post-natal development is not enough to explain the discrepancy in brain weight per gestation day in each group. Rather, to explain the discrepancy, we introduced a deterministic model of cortical neurogenesis, using series summarizing seven neurogenic lineages and based on cell-cycle length, neuroepithelial founder pool size, neurogenic period, and estimates of 
relative progenitor-type population sizes (Table 1). We arrived at two models, based on the analysis of 16 species, that show the highest reliability for predicting cortical neuron numbers in a range of species: a mouse model, which implicates only aRG, IP, and asymmetrically dividing bRG; and a human model, which additionally implicates proliferating progenitors in the SVZ. Each model is defined by the proportional occurrence of each lineage in that model (Table 2). Using the mouse model, with varying proportional occurrences of each lineage, we were able to predict neuron counts within 2$\%$ of the observed counts for mouse and rat, but underestimated neuron counts by more than 80$\%$ in high-GI species (Figure 5; Table S3). Similarly, the human model predicted neuron counts within 5$\%$ for all high-GI species, but overestimated neuron counts by more than 150$\%$ for low-GI species. Increased proportional occurrences of the bRG lineage with increasing brain size was required to achieve estimates with $<$ 
5$\%$ deviation from observed neuron counts in all low-GI species (Table 2; Figure S7). Estimates of proportional occurrences in the mouse, marmoset, and rabbit are supported by previous work detailing relative abundances of different progenitor cell-types during cortical neurogenesis \cite{wang_new_2011,kelava_abundant_2012}, [IK and WBH, \textit{in preparation}]. Evolutionary gain or loss of proliferative potential in the SVZ is an essential mechanistic determinant of neocortical expansion, such that its presence in high-GI species and absence in low-GI species is sufficient and even requisite for explaining neocortical evolution (Figure S8). 

\begin{figure}
\centering
\makebox[\textwidth][c]{
\includegraphics[scale=0.70]{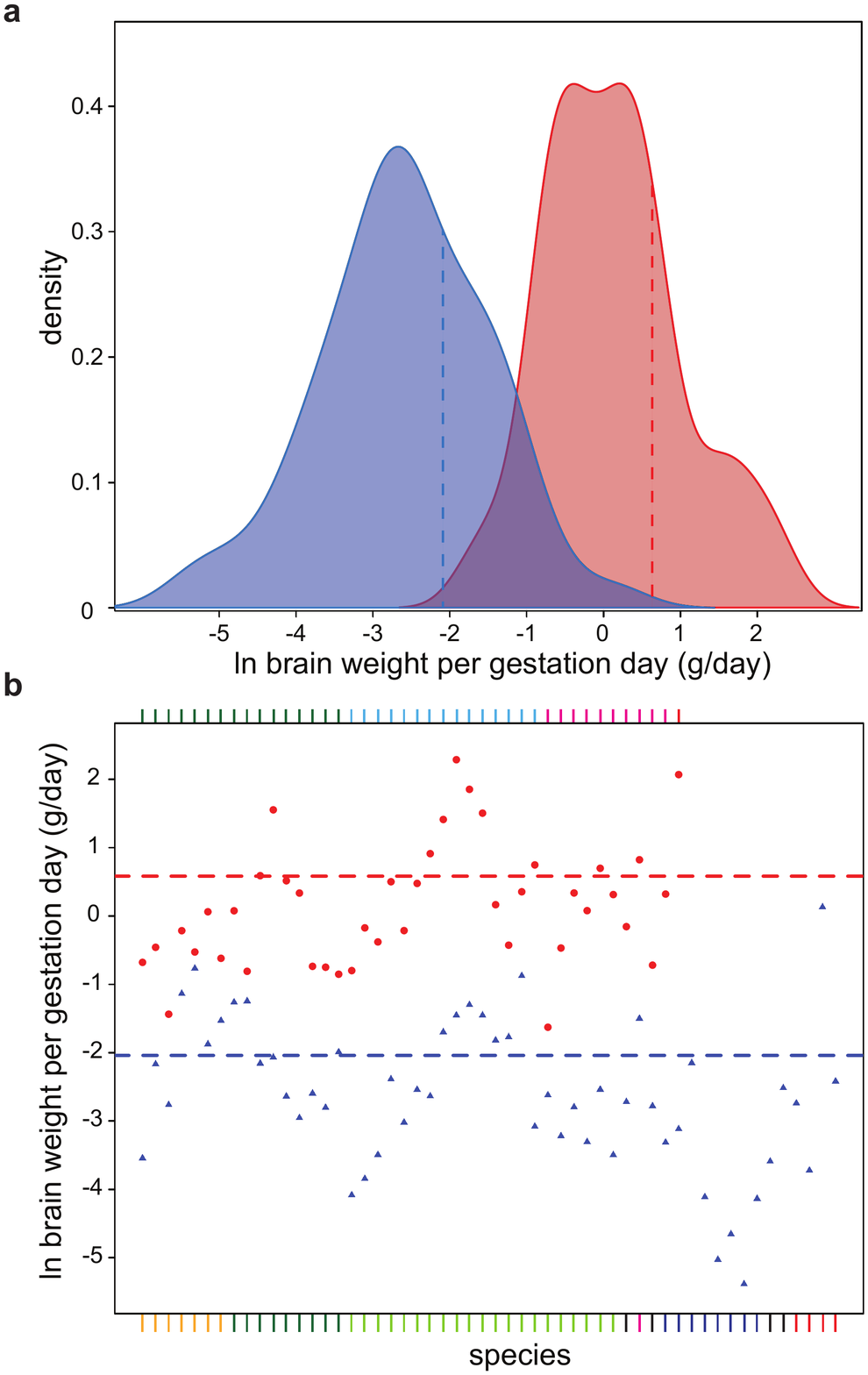}}
\caption{Brain weight per gestation day is considerably greater for high- versus low-GI species. (a) Ln-transformed density plot of brain weight per gestation day for 96 eutherian species listed in Table S1 with GI values of $\leq$ 1.5 (blue) and $>$ 1.5 (red). Note the significantly different means for the two groups (dashed blue and red lines, T = 5.16, d.f. = 41, P = 4 x 10$^{-5}$).(b) Ln-transformed plot of brain weight per gestation day for 96 mammalian species (see a). Dashed blue line, mean value for GI $\leq$ 1.5 (–2.04 $\pm$ 0.047, s.d.); red dashed line, mean value for GI $>$ 1.5 (0.583 $\pm$ 0.050, s.d.). The colors in the index refer to species in Figure 1. See Table S1 for data.}
\label{Figure 4a,b}
\end{figure} 

\section*{Adaptive evolution of proliferative potential in the basal germinal zone}
To simulate the adaptiveness of evolving increased proliferative potential in the SVZ in two lissencephalic species -– mouse and marmoset –- we calculated trade-offs between neuroepithelial founder pool size and neurogenic period using mouse/marmoset and human models of cortical neurogenesis to achieve one billion neurons. We show that, in both species, evolving a lineage of proliferating basal progenitors is between 2- and 6-times more cost-efficient than either expanding founder pool size or lengthening neurogenesis; and that the marmoset, by evolving proliferating progenitors, could keep its observed founder pool size or slightly reduce its neurogenic period to achieve one billion neurons (Figure S9). We further clarified the significance to neuronal output of each progenitor-type with deterministic and stochastic models of temporal dynamics and progenitor cell-type variables. From these we conclude that basal progenitors are increasingly necessary in larger brains and that achieving 10$^9$ neurons is 
statistically implausible in the absence of proliferative basal progenitors (Table S4). Finally, we described the dynamics of asymmetric \textit{versus} symmetric progenitors, isolated from their observed lineage beginning at the apical surface, by introducing three ordinary differential equations (ODEs) modeling a self-renewing cell that generates either a differentiated cell or proliferative cell. The ODEs describe a self-renewing mother progenitor, which can generate either a neuron or a proliferative daughter at each division. The proliferative daughter is allowed one proliferative division followed by self-consumption. The likelihood of a neuron or proliferative daughter being generated by the mother, therefore, is interdependent. We also include the pool of mother progenitors as a linear variable. We show that neuronal output of the system increases dramatically when both the initial pool of self-renewing cells and the likelihood of those initial cells to generate proliferative, rather than 
differentiated, cells approaches saturation (Figure S10).

\begin{figure}
\centering
\makebox[\textwidth][c]{
\includegraphics[scale=0.7]{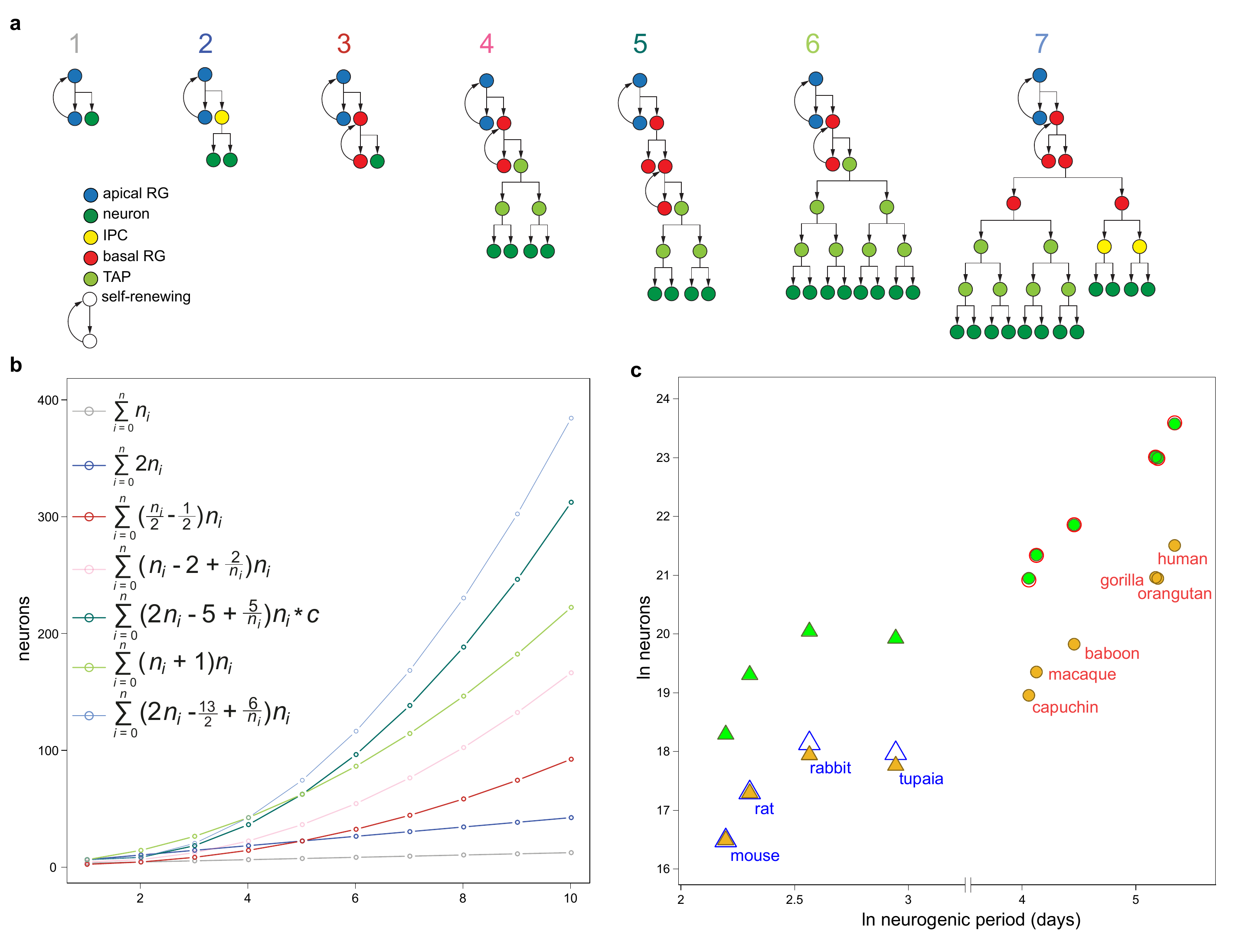}}
\caption{Distinct combinations of progenitor lineages are required to predict neocortical neuron numbers for low- \textit{versus} high-GI species. (a) Schematics of the 7 lineages used to construct neurogenic output in species. (b) Plotted neuronal output of the lineages in (a) beginning with two cells, over 10 divisions. Series in the legend summarize the neuronal output of each lineage, where $n_i$ is the number of \textit{i} divisions. A constant, c = 0.989, is incorporated into the series for lineage 5, allowing the series to converge on the true value of the lineage output as the number of divisions becomes increasingly numerous. (c) Ln-transformed plot of observed neuron counts as a function of neurogenic period for 4 species with a GI $\leq$ 1.5 (open blue triangles) and 6 species with a GI $>$ 1.5 (open red circles). Predicted neuron counts were calculated using combinations of the lineages in (a) that accurately fitted to the observed neuron counts either for mouse (closed gold symbols) or human (
closed green symbols). Note that the mouse model implicates only lineages 1--3; the human model only lineages 2--7; and that lineages 4--7 were considered interdependent, such that an increase (or decrease) in the occurrence of one of these lineages necessitated an attendant increase (or decrease) in the others. See Table 1 for observed and predicted neuron counts and Table 2 for the proportional contribution of each lineage for mouse and human.}
\label{Figure 5}
\end{figure} 

\section*{Discussion}
The emergence of new structures, in the most general sense, is typically limited to selection on existing developmental processes; and conserved pathways may persist, over evolutionary time, even when the phenotype is transformed or unexpressed \cite{mayr_emegergence_1960,shubin_fossils_1997,hall_descent_2003}. However, it is also evident that development may be adapted without affecting phenotype (e.g., \cite{bolker_comparison_1994,kalinka_evolution_2012}). Therefore, in order to understand selective pressures acting on a discontinuous or convergent trait, it is necessary to investigate the underlying developmental processes generating it. We have shown that a gyrencephalic neocortex is ancestral to mammals, which is concordant with evidence \cite{romiguier_genomic_2013} that the mammalian ancestor was large ( $>$ 1kg) and long-lived ( $>$ 25-year lifespan) and, furthermore, provides considerable resolution to recent evidence for a gyrencephalic eutherian ancestor \cite{oleary_placental_2013} by sampling 
nearly twice as many species and categorizing gyrencephaly as a continuous, rather than a binary, trait.  More surprisingly, we show that convergent evolution of higher-orders of gyrencephaly along divergent lineages has been accompanied by two distinct constellations of physiological and life-history paradigms. Specifically, species with a GI $>$ 1.5, which is commensurate with one billion cortical neurons, exhibit patterns of development and life-history that are distinct from species with a GI $<$ 1.5, irrespective of phylogeny. This implies that there is a considerable constraint on either the ability of species of a given neocortical size to exploit certain ecologies or the potential for species of a given ecology to freely adapt neocortical size. Even marine mammals, whose selection pressures are \textit{sui generis}, may largely be held to the same evolutionary stereotyping as terrestrial mammals (Figure S11). Furthermore, no species –- with the exception of the house cat (\textit{Felis catus}), which 
may be under unique selection pressures due to its ten-thousand-year-old domestication \cite{driscoll_taming_2009} –- falls within the limits of the GI or neuronal threshold range (Figure 3d). While our results countenance previous studies showing associations between physiological and life-history traits in mammals (see \cite{martin_problems_2005}), we identify those traits to have a bimodal distribution, rather than to vary allometrically, across species. This distribution depicts a Waddington-type landscape for neocortical expansion -- albeit relevant at the species-level -– wherein the threshold represents an adaptive peak requiring a particular adaptation in neurogenic programming within a population for traversal. Our results may explain this landscape by mechanistic differences occurring during cortical neurogenesis between species above and below the threshold: the necessity of proliferative basal progenitors in high-GI species and their putative absence in low-GI species. The adaptation of 
proliferative basal progenitors may be tantamount to a relaxation of constraints along lineages leading to larger-brained species \cite{boddy_comparative_2012}. Furthermore, our human model clearly shows that the same neurogenic lineages in the same proportions are required to generate the neocortices of monkeys, apes, and humans, and may even be extended to carnivores, cetartiodactlys, and other high-GI species (Figure S11), demonstrating that neurogenic period alone may be sufficient to explain differences in neocortical size between any species in the same GI group (Figure S12). If differences in neurogenesis among high-GI species can be largely explained by variation in neurodevelopmental timing, we may expect conservation at the genomic level in regions regulating that timing \cite{lewitus_neocortical_2013}.

We propose that proliferative basal progenitors, rather than simply an abundance of asymmetrically dividing bRGs in an expanded SVZ, are necessary and sufficient for the evolution of an expanded and highly folded neocortex in mammals. We conclude that an increase in proliferative potential in the basal neurogenic program is an adaptive requirement for traversing an evolutionary threshold. But because we reconstruct the eutherian ancestor to have a GI value of 1.48 $\pm$ 0.13 (s.e.m.), which falls within the range of the observed threshold, we are left with an ambivalent evolutionary history for mammalian neocortical expansion: either (i) proliferative basal progenitors are ancestral to all eutherian mammals and were selected against along multiple lineages (e.g., rodents, strepsirrhines), so that the ultimate loss of basal proliferative potential in certain taxa, and therefore the evolution of low-GI species, is the result of divergent developmental adaptations; or (ii) proliferative basal progenitors are 
not ancestral to eutherian mammals, but evolved convergently along multiple lineages, in which case the developmental process for their inclusion in neurogenic programming may be conserved, even if that process was unexpressed for long stretches of mammalian evolution. While both of these histories are speculative, we have nonetheless revealed an important insight into mammalian evolution: a threshold exists in mammalian brain evolution; neocortical expansion beyond that threshold requires a specific class of progenitor cell-type; and the difference in neurogenic programming between any species on the same side of that threshold does not require novel progenitor-types or adaptations in progenitor-type behavior. Further research into the conservation of genomic regions regulating the expression of proliferative basal progenitors, either at the ventricle or through maintenance of a proliferative niche in the SVZ, in low- \textit{versus} high-GI species may be sufficient to determine whether the mechanism for 
neocortical expansion has evolved independently in distantly related species or is the product of a deep homology in mammalian neurogenesis. 

\section*{Materials and Methods}
\subsection*{Calculating GI}
We calculated GI using images of Nissl-stained coronal sections from http://brainmuseum.org. We used 10-22 sections, equally spaced along the anterior-posterior axis of the brain, for each species (Figure S1). The inner and outer counters of the left hemisphere were traced in Fiji (http://fiji.sc/wiki/index.php/Fiji). The values calculated are marked with an asterisk in Table S1. Additional GI values were collected from the literature (Table S1; External Database 1). Species (e.g., platypus) whose cortical folding has been described \cite{goffinet_what_2006,rowe_organisation_1990}, but not measured according to the method established in \cite{zilles_human_1988}, were omitted from our analyses (see \textit{Reconstructing the evolutionary history of GI}). Work in humans and baboons has shown that inter-indvidual variation in GI is not enough to outweigh interspecific differences \cite{rogers_genetic_2010,toro_brain_2008}.  

\subsection*{Stochastic mapping of GI across the mammalian phylogeny}
We used a comprehensive phylogenetic approach to map 41 life-history and physiological character traits collected from the literature (Tables S1,S2) onto hypotheses of phylogenetic relationships in Mammalia, in order to examine how those traits correlate, over evolutionary time, with degree of gyrencephaly. Continuous character traits were discretized using the consensus of natural distribution breaks calculated with a Jenks-Caspall algorithm \cite{jenks_error_1971}, model-based clustering according to the Schwarz criterion \cite{fraley_model-based_2002}, and hierarchical clustering \cite{szekely_hierarchical_2005}. Character histories were then corrected for body mass with a phylogenetic size correction \cite{collar_discordance_2006} and summarized across the phylogeny using posterior probabilities. Associations between individual states of each character trait along those phylogenetic histories were calculated in SIMMAP (v1.5) using empirical priors \cite{bollback_simmap:_2006}; the association 
between any two states was a measure of the frequency of occurrence (i.e., the amount of branch length across the tree) of those states on the phylogeny. The sums, rates, and types of changes for GI and body weight were plotted as mutational maps to assess directional biases in their evolution \cite{cunningham_limitations_1999,huelsenbeck_detecting_2003,lewitus_life-history_2011}. The phylogeny used in this analysis was derived from a species-level supertree \cite{bininda-emonds_delayed_2007}. We appreciate that the phylogenetic hypothesis reconstructed by \cite{meredith_impacts_2011} gives notably deeper divergence dates for mammalian subclasses, however, not enough of our sampled species were included in this reconstruction for it to be useful here.

\subsection*{Reconstructing the evolutionary history of GI}
Variation in the mode and tempo of a continuous character trait is not always best characterized by a random walk (i.e., Brownian motion). Therefore, we compared a range of evolutionary models on the phylogenetic distribution of GI to find the best fit for the data \cite{felsenstein_maximum-likelihood_1973,harmon_geiger:_2008,omeara_testing_2006,paradis_ape:_2004}. Log-likelihood scores for each model were tried against the random walk score using the cumulative distribution function of the $\chi$$^2$ distribution. Maximum-likelihood ancestral character states of GI and rate-shifts in the evolution of GI were then constructed using the best-fit model, with the standard error and confidence intervals calculated from root node reconstruction in PDAP using independent contrasts \cite{garland_jr_using_2000,garland_phylogenetic_2005,maddison_mesquite:_2011}. Although a number of putatively lissencephalic non-eutherians were unavailable for our analyses (see \textit{Calculating GI}), we nonetheless reconstructed 
alternative ancestral GI values that included one hypothetical monotreme and three hypothetical marsupials (Table S5). To trace evolutionary changes in GI at individual nodes and along lineages, we used a two-rate mode that highlighted the differences in high ($>$ 1) versus low ($<$ 1) root-to-tip substitutions and then sampled rates based on posterior probabilities across the tree using a Monte Carlo Markov Chain. We assumed that transitioning between adjacent GI values had the highest likelihood of occurrence. The rate at a given node could then be compared to the rate at the subsequent node to determine if a rate transition was likely. We corroborated these results using the \textit{auteur} package \cite{eastman_novel_2011}, which calculates rate-transitions at internal nodes under the assumption of an Ornstein-Uhlenbeck selection model \cite{butler_phylogenetic_2004} over one million Monte Carlo sampling iterations drawn from random samplings of posterior distributions of lineage-specific rates. 
Scaling relationships were determined for GI as a function of all continuous life-history and physiological traits, including adult cortical neuron counts. For three insectivore (\textit{Sorex fumeus, Blarina brevicauda, Scalopus aquaticus}) species, data were available for neuron counts but not GI, and therefore we extrapolated the GI of those species based on their closest phylogenetic relatives. Finally, to test whether the bimodal distribution of GI may be influenced by the topology of the mammalian phylogenetic tree, we used an expectation-maximization algorithm. Each simulated trait was given the same variance as GI (Figure S5) and the result was averaged over 10$^4$ simulated datasets. None of the simulations produced the same bimodal distribution of species observed for GI data. 

\subsection*{Estimating neuroepithelial founder pool populations}
We estimated neuroepithelial founder pool populations for mouse and human. For the mouse, we used coronal sections of an E11.5 mouse embryo obtained from the Allen Brain Atlas \cite{lein_genome-wide_2007}. We obtained 19 sections equidistantly spaced along the anterior-posterior axis of the brain. The length of the ventricular surface of the dorsal telencephalon was manually traced in Fiji \cite{schindelin_fiji:_2012} on each section starting from the point above the nascent hippocampus and ending in the point above the lateral ganglionic eminence. The horizontal length of the embryonic brain at E11.5 was measured with images from \cite{bejerano_distal_2006}. Using the coronal and horizontal measurements, we constructed a polygon representing the ventricular surface of the dorsal telencephalon and calculated the area of this surface in Fiji. We measured the surface area of the end-feet of neuroepithelial cells using EM images of the coronally cut apical surface of an E11.5 embryonic brain (Table S6). The 
diameter of a single cell was calculated by measuring the distance between the adherens junctions. We corroborated these end-feet calculations with published immunofluorescence stainings of the apical complex (ZO1 and N-cadherin) from an \textit{en face} perspective \cite{bultje_mammalian_2009,marthiens_adherens_2009}. The average surface area of a single end-foot was calculated by approximating the end-foot as a hexagon; and the number of founder cells was estimated by dividing the surface of the dorsal telencephalon by the surface of an individual end-foot of the neuroepithelial cell, such that
\\
\begin{equation}
\frac{Surface\ area (\mu m^2)}{2\pi (\frac{1}{2} Endfoot\ diameter (\mu m^2)}\frac{\sqrt{3}}{2} = founders
\end{equation}
\\
Our final mouse values were comparable to those previously published \cite{haydar_role_2000}. For the human, we followed the same procedure, using 10 coronal sections and one horizontal section of a gestation week (GW) 9 brain \cite{bayer_atlas_2006}. End-feet were calculated using EM images of the apical surface of a human brain at GW13. The measurements are available in Table S6. 
Because the number of founder cells per surface area was nearly equivalent in mouse and human (~ 4 x 10$^5$/mm$^2$), we used this ratio, along with data on ventricular volume collected from the literature (Table S1; Table S2; External Database 1), to estimate neuroepithelial founder cell populations for a further 14 species (Table 1). For species where no data on ventricular volume were available, values were estimated based on a regression analysis against brain weight (Figure S6). Ventricular volume was then converted to surface area for each species by approximating the ventricle as a cylinder with a 4.5-to-1 height-to-diameter proportion. Ventricular volume-derived ventricular surface area estimates were corroborated with the surface areas calculated from the literature for mouse and human. Founder cell estimates were then computed based on the densities derived above for mouse and human. Using this method, but alternately ignoring our mouse and human calculations to define the parameters, we were able 
to predict mouse and human values within 10$\%$ of our calculations, respectively.

\subsection*{Mathematical modeling of neurogenesis}
Workers have demonstrated the occurrence of three primary lineages of neuronal generation in mouse neurogenesis \cite{fietz_cortical_2011} and a further two lineages in human neurogenesis \cite{hansen_neurogenic_2010}. While there is evidence for at least one additional lineage in mouse \cite{noctor_cortical_2004}, and further lineages may be speculated, we limited our model to the five that are considered to contribute most significantly to neuronal output \cite{rakic_evolution_2009,lui_development_2011,molnar_evolution_2011}. The sequence of neuron generation in each of these five lineages was summarized in series and solved numerically (Figure 5b). Neurogenic period was either taken from the literature (External Database 1) or estimated based on a regression analysis of neurogenic period as a function of gestation period (Figure S6). Neurogenic period in human was estimated using empirical observations from the literature \cite{bystron_first_2006,howard_cortical_2006,malik_neurogenesis_2013}. The 
averaged cell-cycle length for apical and basal progenitors from the mouse (18.5 hours) was used for all non-primates (\cite{arai_neural_2011}; Figure S13); averaged cell-cycle length for cortical areas 17 and 18 from the macaque (45 hours) was used for catarrhines \cite{lukaszewicz_g1_2005,betizeau_accepted}; and an intermediary cell-cycle length (30 hours), based on personal observations in marmoset, was used for platyrrhines. Diminishing numbers of neuroepithelial cells have been observed to continue to proliferate at the ventricle until E18.5 in the mouse \cite{haubensak_neurons_2004}. Therefore, final neuroepithelial founder pool estimates were calculated from the aforementioned by evenly decreasing the value of $\alpha$ in the Sherley equation \cite{sherley_quantitative_1995} from 1 at E9.5 to 0 at E18.5 in the mouse and at comparable neurogenic stages in other species. Neuron numbers were calculated for each species from combinations of lineages. The proportional contribution of each lineage for 
each species was parameterized according to existing data on progenitor cell-type abundances in mouse \cite{wang_new_2011}, marmoset \cite{kelava_abundant_2012}, and rabbit [IK and WBH, \textit{in preparation}]. Where no such data were available, proportional contributions were permutated for all lineages until a best-fit estimate, based on cortical neuron numbers taken from the literature \cite{azevedo_equal_2009,gabi_cellular_2010,herculano-houzel_coordinated_2010,herculano-houzel_not_2011}, was achieved (Tables 1,2). Each lineage was assumed to occur from the first to final day of neurogenesis, although this is only approximately accurate. Finally, because of published estimates of postnatal apoptosis in the mammalian cortex \cite{burek_programmed_1996,hutchins_why_1998,bandeira_changing_2009}, we assumed neuron counts to be 1.5-fold higher at the termination of neurogenesis than in the adult brain; therefore, neuron number at the termination of neurogenesis was estimated in each species by 
multiplying neuron numbers collected from the literature by 1.5. This multiplication is not represented in Table 1.

\subsection*{Calculating the effects of proliferative progenitors on neuronal \\output}
Trade-offs in adapting a human lineage combination with either an expanding neuroepithelial founder pool or lengthening neurogenic period were tested for the mouse (Mus musculus) and marmoset (\textit{Callithrix jacchus}), two lissencephalic species whose cell-type proportions during neurogenesis have been documented \cite{noctor_cortical_2004,wang_new_2011,kelava_abundant_2012}. To estimate the relative reproductive value and stable-stage proportions of each of the lineages in the mouse and human models, we constructed a stage-structured Lefkovitch matrix, using sums of the lineage series (after 100 cycles) as fecundity values and complete permutations of the proportional contributions of each lineage as mortality values. The altered growth-rates of each lineage were calculated by excluding lineages one at a time and assuming 100$\%$ survival in the remaining lineages. We introduced three ODEs to explore the average dynamics of asymmetric \textit{versus} symmetric progenitors, such that: if a(t), b(t), and 
c(t) are the numbers of asymmetrically dividing cells, differentiated cells, and proliferative cells, respectively, then,

\begin{equation}
\centering
\frac{da}{dt}=0 
\end{equation}
\begin{equation}
\centering
\frac{db}{dt} = ra + 2rc 
\end{equation}
\begin{equation}
\centering
\frac{dc}{dt} = (1-r)a + (1-2r)c
\end{equation}
\\
where \textit{r} is equal to growth-rate. If \textit{a(t)}=a$_0$, then
\\
\begin{equation}
\centering
b(t) = \frac{2r}{1-2r}(c_0 + \frac{1-r}{1-2r}a_0)(e^{(1-2r)t}-1)-\frac{ra_0}{1-2r}t+b_0
\end{equation}
and
\begin{equation}
\centering
c(t) = (c_0 + \frac{1-r}{1-2r}a_0)e^{(1-2r)t}-\frac{ra_0}{1-2r}a_0
\end{equation}
\\
We calculated the effect on neuronal output of increasing the likelihood of symmetrically dividing daughter progenitors in the lineage (Figure S10). The interdependent growth-rates in the model reflect a purely mechanistic interpretation of determining neuronal output from a finite pool of asymmetrically dividing cells. The ODEs, therefore, may not reflect differential regulation of neuronal output via direct \textit{versus} indirect neurogenesis. The daughter proliferative cells are designed to carry out one round of proliferation followed by a final round of self-consumption.\\

\noindent \textbf{Acknowledgments}: We would like to thank Michaela Wilsch-Bräuninger for providing electron microscopy data; and Michael Hiller, Yannis Kalaidzidis, Fong Kuan Wong, and Alex Sykes for helpful comments on the manuscript; and EL would like to thank Evan Charles for helpful discussion. IK was a member of the International Max Planck Research School for Molecular Cell Biology and Bioengineering and a doctoral student at the Technische Universitat Dresden. 
\begin{figure}[!h]
\centering
\makebox[\textwidth][c]{
\includegraphics[scale=0.80]{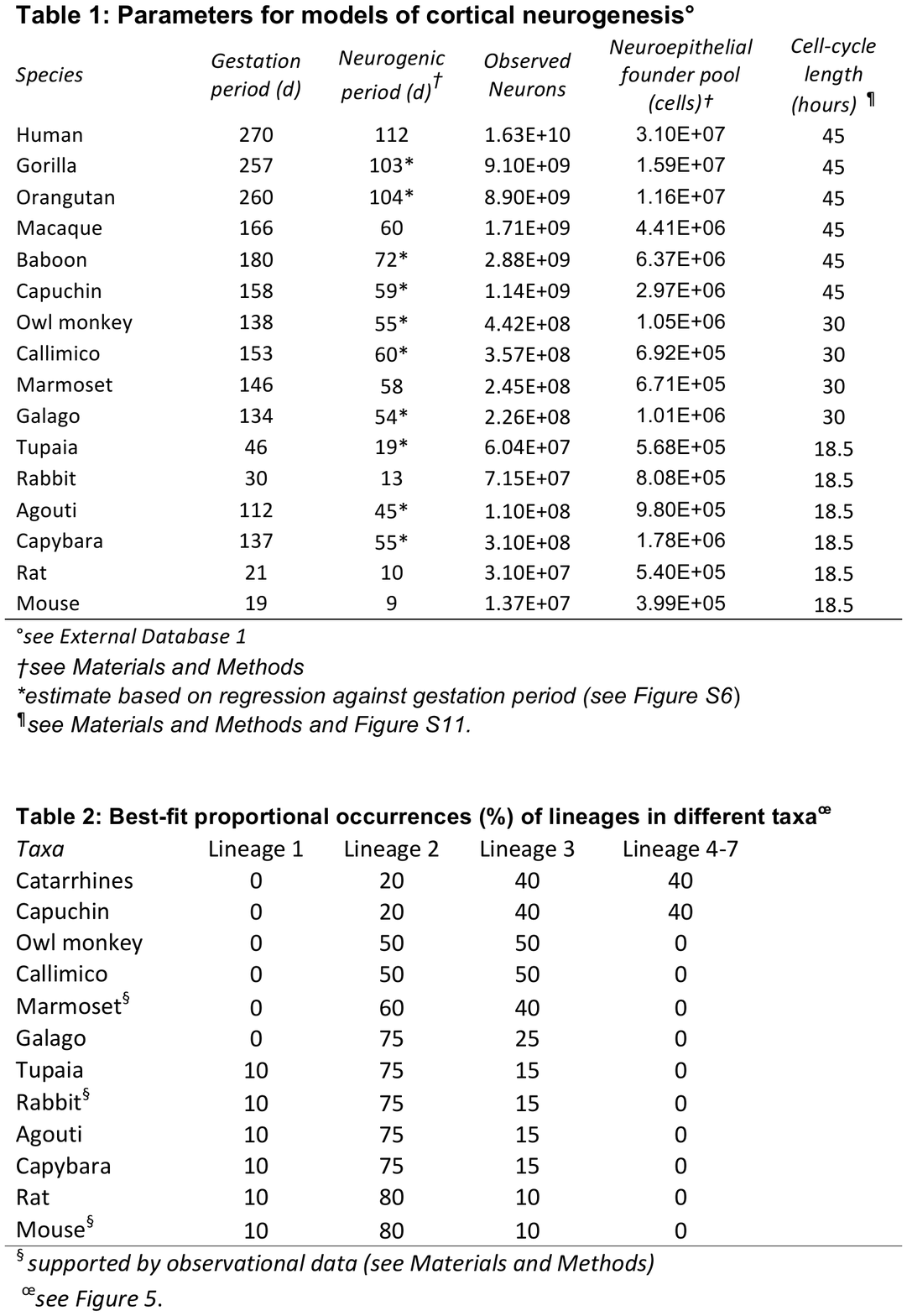} }
\label{Table1}
\end{figure}
\setstretch{1.1}
\pagebreak
\\
\bibliography{50eggs}

\pagebreak
\setcounter{figure}{0}
\makeatletter
\renewcommand{\thefigure}{S\@arabic\c@figure}
\makeatother
\begin{figure}[hb]
\centering
\makebox[\textwidth][c]{
\includegraphics[scale=.8]{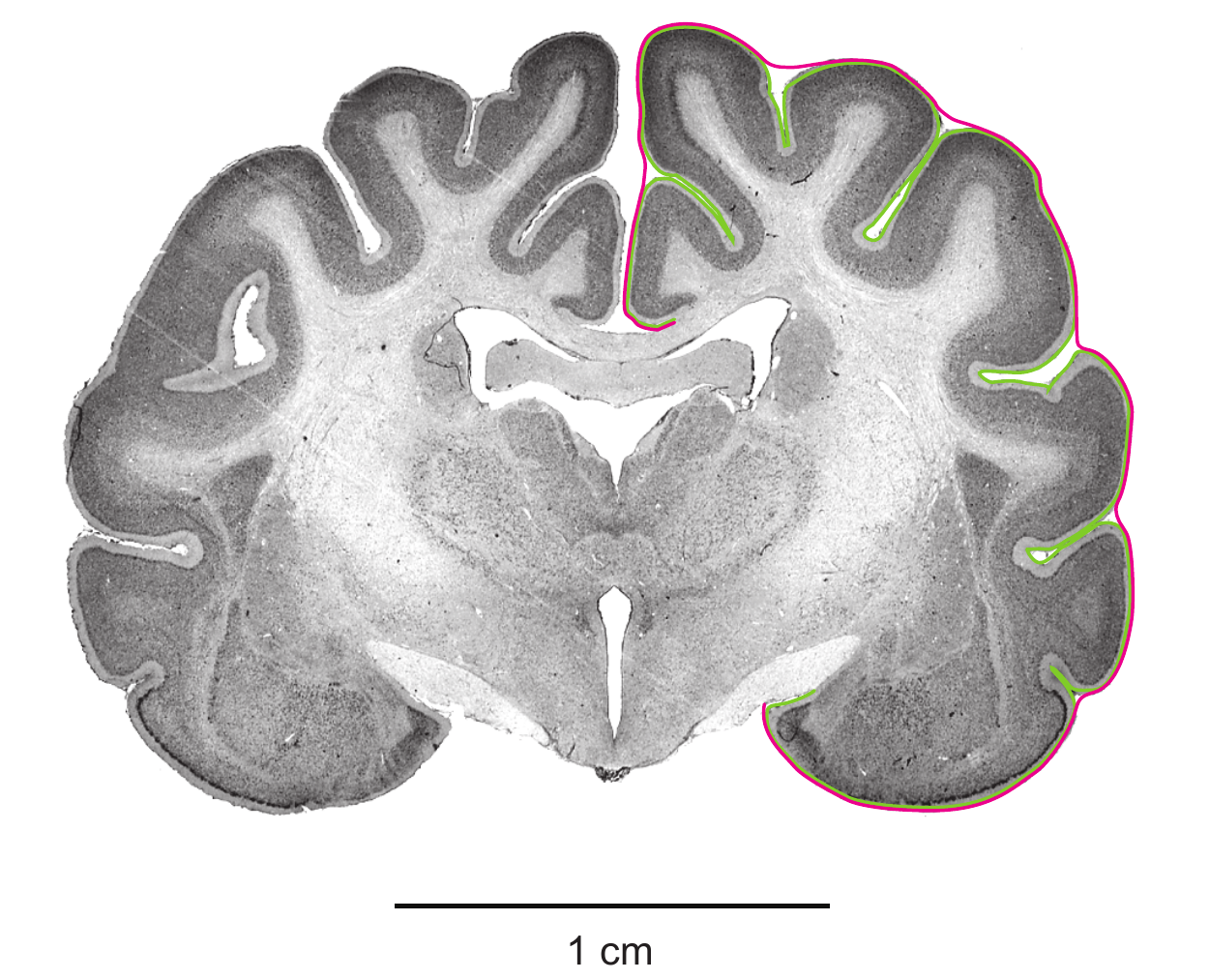}}
\caption*{\textbf{Figure S1:} Coronal section of the brain of an adult house cat (\textit{Felis catus}) (obtained from www.brainmuseum.org) illustrating the method used to calculate GI values as described in \cite{zilles_human_1988}. Green line, actual contour; magenta line, hypothetical ‘outer’ contour. }
\label{Figure S1}
\end{figure} 
\pagebreak
\begin{figure}[hb]
\centering
\makebox[\textwidth][c]{
\includegraphics[scale=0.80]{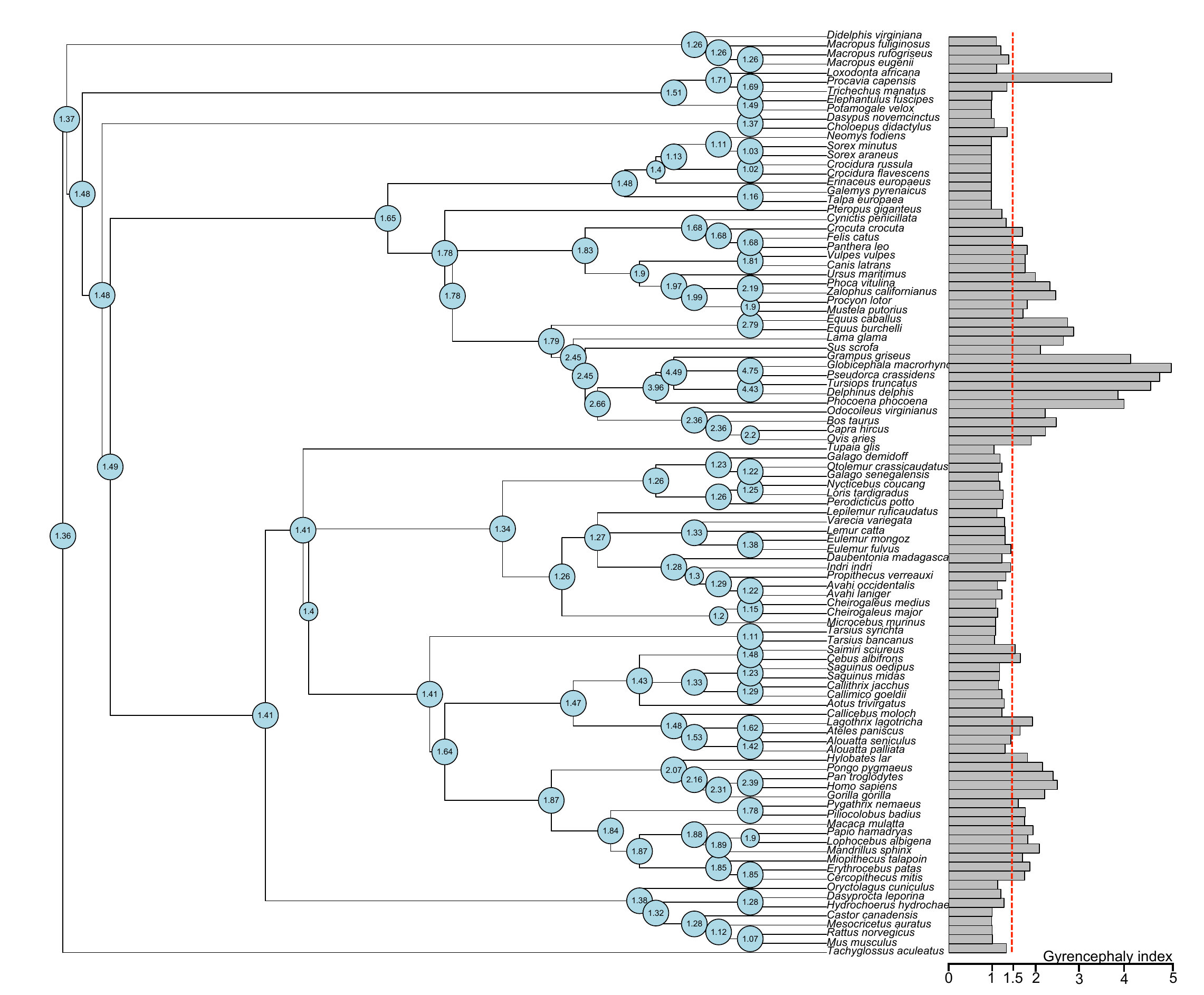}}
\caption*{\textbf{Figure S2:}Maximum-likelihood ancestral node reconstruction of GI values at all internal nodes based on a delta ($\delta$ = 2.635) selection model. Barplot shows the distribution of GI values across the phylogeny; dashed red line indicates GI = 1.5.}
\label{Figure S2}
\end{figure} 

\begin{figure}[hb]
\centering
\makebox[\textwidth][c]{
\includegraphics[scale=0.55]{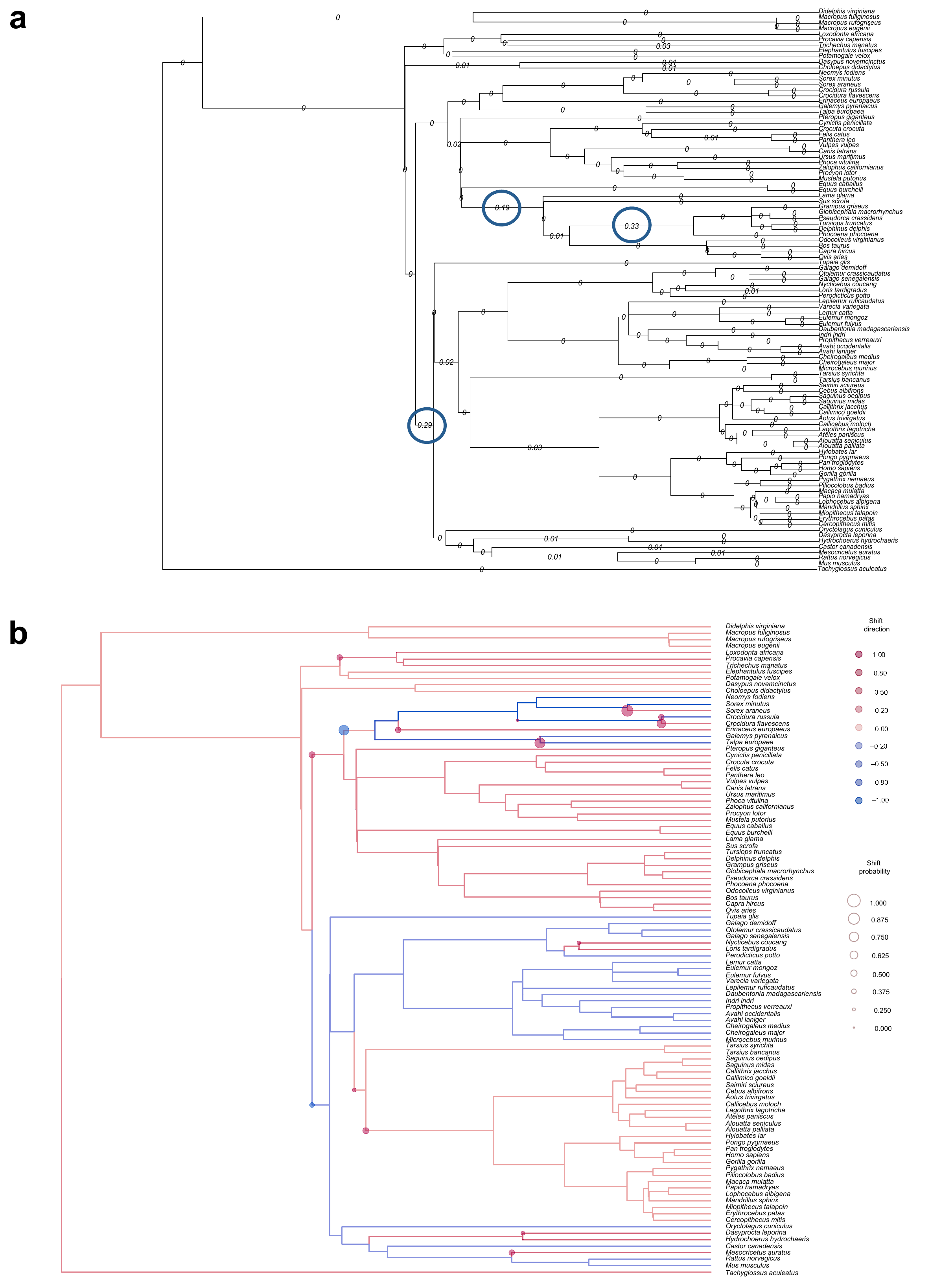}}
\caption*{\textbf{Figure S3:} Rate-transitions in the mutation rate of GI values along lineages of the mammalian phylogeny. (a) A two-mode selection model that weights low over high root-to-tip substitutions. Numbers on the branches indicate the change in mutation-rate compared to the previous branch; 0 values indicate no significant change, values $>$ 0 indicate significant change (P $<$ 0.05). Note the especially high rate-transitions leading to primates, cetartiodactyls, and cetaceans (open blue circles). (b) Mutation- and transition-rate estimates of GI values using an Ornstein-Uhlenbeck selection model. Branches are colored to illustrate whether the mutation-rate estimates along each lineage are above (red) or below (blue) the median rate (orange); nodes are circled to indicate the posterior support of a transition-rate-shift event. The gradient of colors (see key) indicates the degree of deviation of the mutation-rate estimates (branches) and transition-rate estimates (nodes) from the median, with the 
highest deviation being arbitrarily set to $\pm$ 1.0 and the median to 0.0; the size of the circles (see key) at the nodes indicates the degree of posterior support for a transition-rate-shift event, with the highest value being arbitrarily set to 1.0 and lack of support to 0.0. Note that simians have evolved GI values at a rate consistent with the mammalian median.}
\label{Figure S3}
\end{figure}

\begin{figure}[hb]
\centering
\makebox[\textwidth][c]{
\includegraphics[scale=.75]{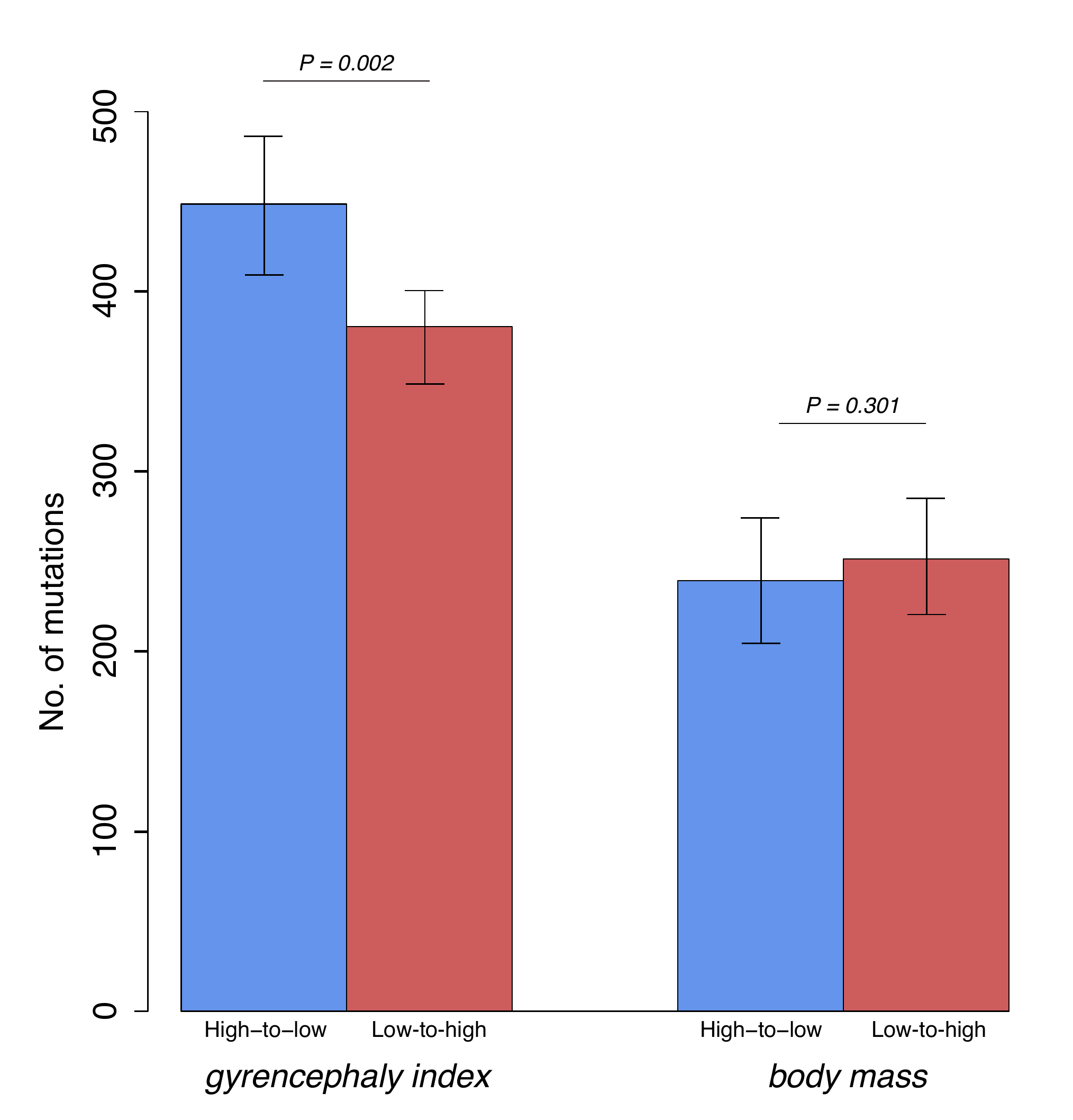}}
\caption*{\textbf{Figure S4:} Barplots of types of transitions over mammalian evolution between four GI groups (see Figure 2a) and between five body mass groups averaged over 105 simulations. The number of total transitions from one GI or body mass group to another is summed as either high-to-low or low-to-high transitions. Note that significantly more high-to-low than low-to-high transitions are observed for GI, but that no significant difference in type of transition is observed for body mass.}
\label{Figure S4}
\end{figure} 

\begin{figure}[hb]
\centering
\makebox[\textwidth][c]{
\includegraphics[scale=1.6]{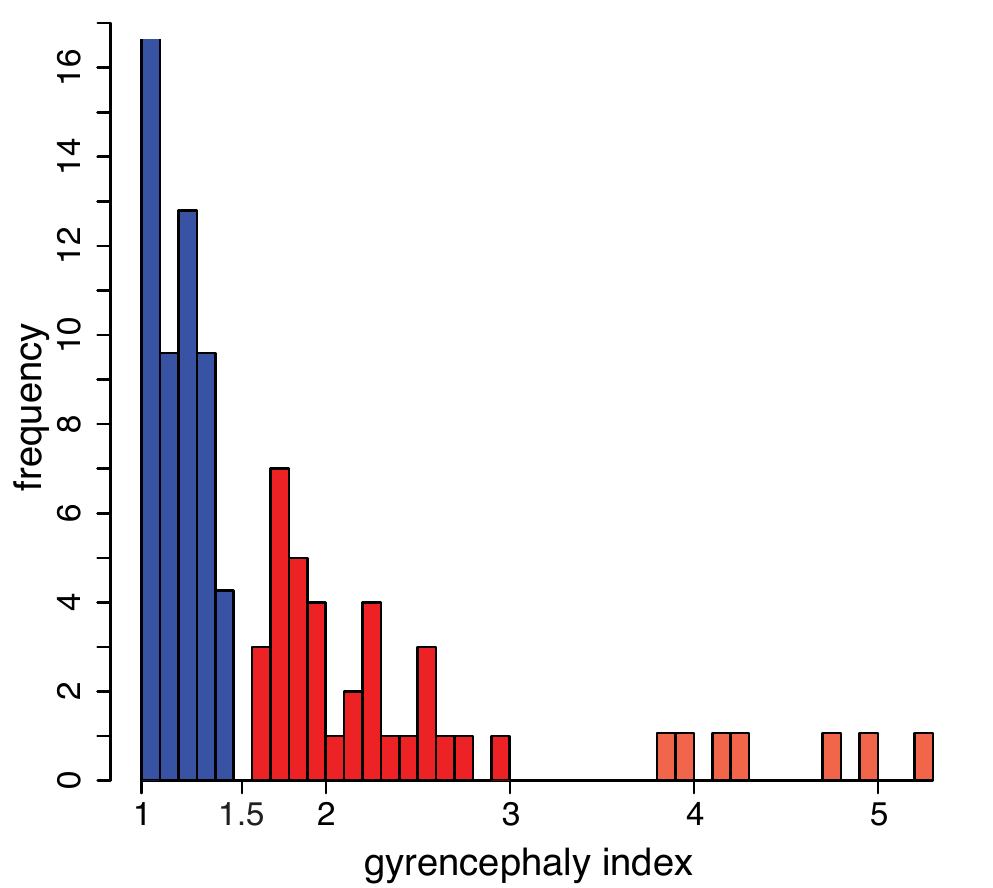}}
\caption*{\textbf{Figure S5:} The bimodal distribution of GI values across the phylogeny is non-random. A histogram showing the frequency of occurrence of GI values, binned at 0.05 intervals, for the 102 mammalian species listed in Table S1. Blue, GI values $\leq$ 1.5; red, GI values $>$ 1.5. The bimodal distribution of GI values shows a natural break at GI = 1.5, which is supported by energy-based hierarchical clustering (see Figure 2b). Note the possibility for a third GI group (GI $>$ 3, tomato red), constituting cetaceans and elephant; however, we have too few sampled species from these orders to assess the group decisively (see Figure S11).}
\label{Figure S5}
\end{figure} 

\begin{figure}[hb]
\centering
\makebox[\textwidth][c]{
\includegraphics[scale=0.50]{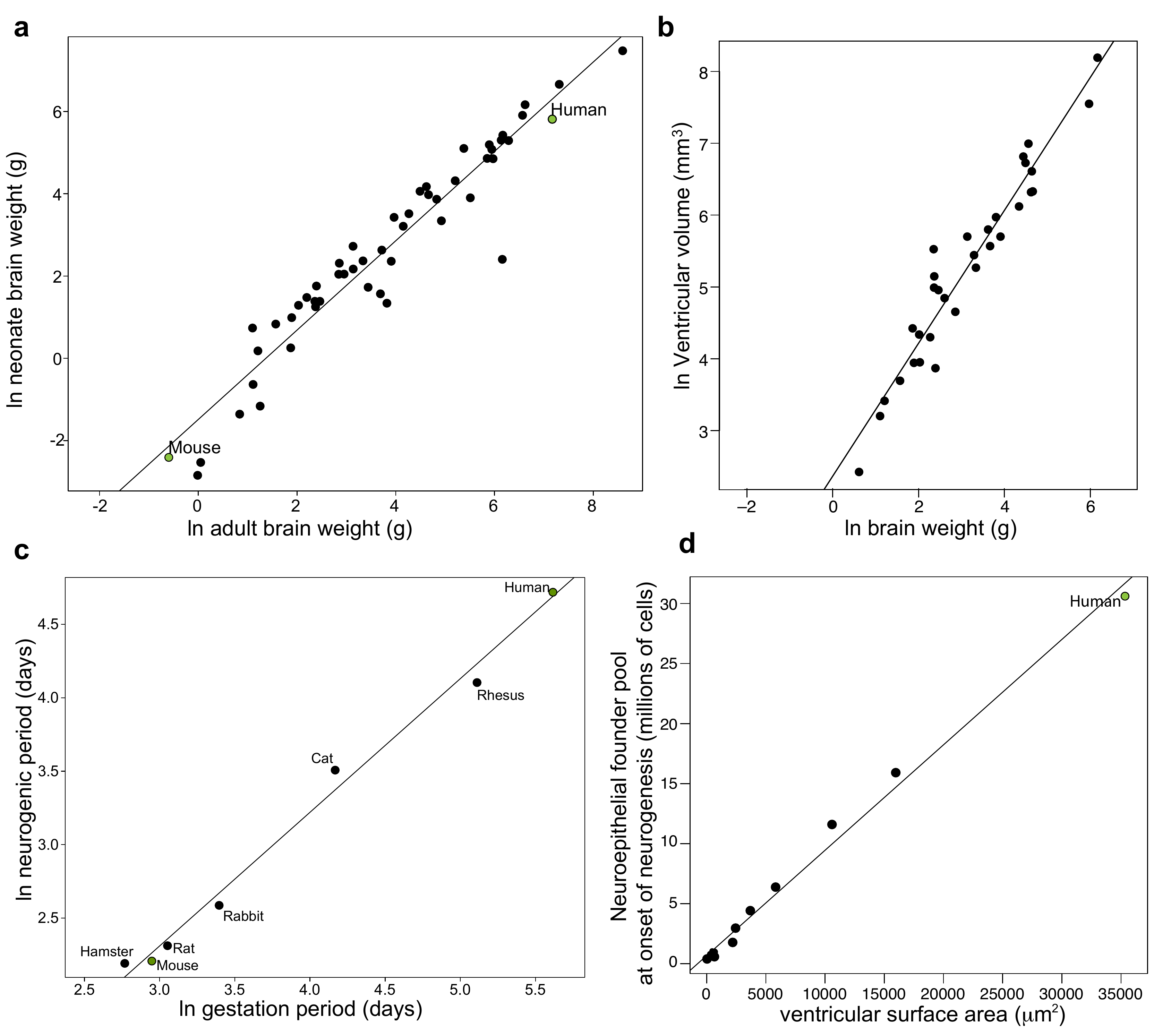}}
\caption*{\textbf{Figure S6:} Ln-transformed plots of neonate brain weight (a) and ventricular volume (b) as functions of adult brain weight, neurogenic period as a function of gestation period (c); and a plot of neuroepithelial founder cells as a function of ventricular surface area (d). (a) Neonate brain weight scales linearly with adult brain weight for 52 eutherian species (y = 1.09x –- 1.49, R$^2$ = 0.92, P = 6 x 10$^{-7}$).
(b) Ventricular volume scales linearly with adult brain weight for 30 eutherian species (y = 0.93x + 2.37, R$^2$ = 0.93, P = 9 x 10$^{-8}$). 
(c) Neurogenic period scales linearly with gestation period for a sample of six species (y = 0.91x –- 0.42, R$^2$ = 0.94, P = 0.0002), spanning two mammalian superorders. Predicted neurogenic period is shown for human.
(d) Ventricular surface area, converted from ventricular volume (see Methods), scales linearly with our estimated neuroepithelial founder populations (y = 6.7 x 10$^5$ + 878x, R$^2$ = 0.94, P = 5 x 10$^{-8}$).
(a, c) Note that these plots demonstrate the strong predictive powers of adult brain weight and gestation period for neonate brain weight and neurogenic period, respectively, validating the assumptions made in Figure 4.}
\label{Figure S6}
\end{figure} 

\begin{figure}[hb]
\centering
\makebox[\textwidth][c]{
\includegraphics[scale=1.15]{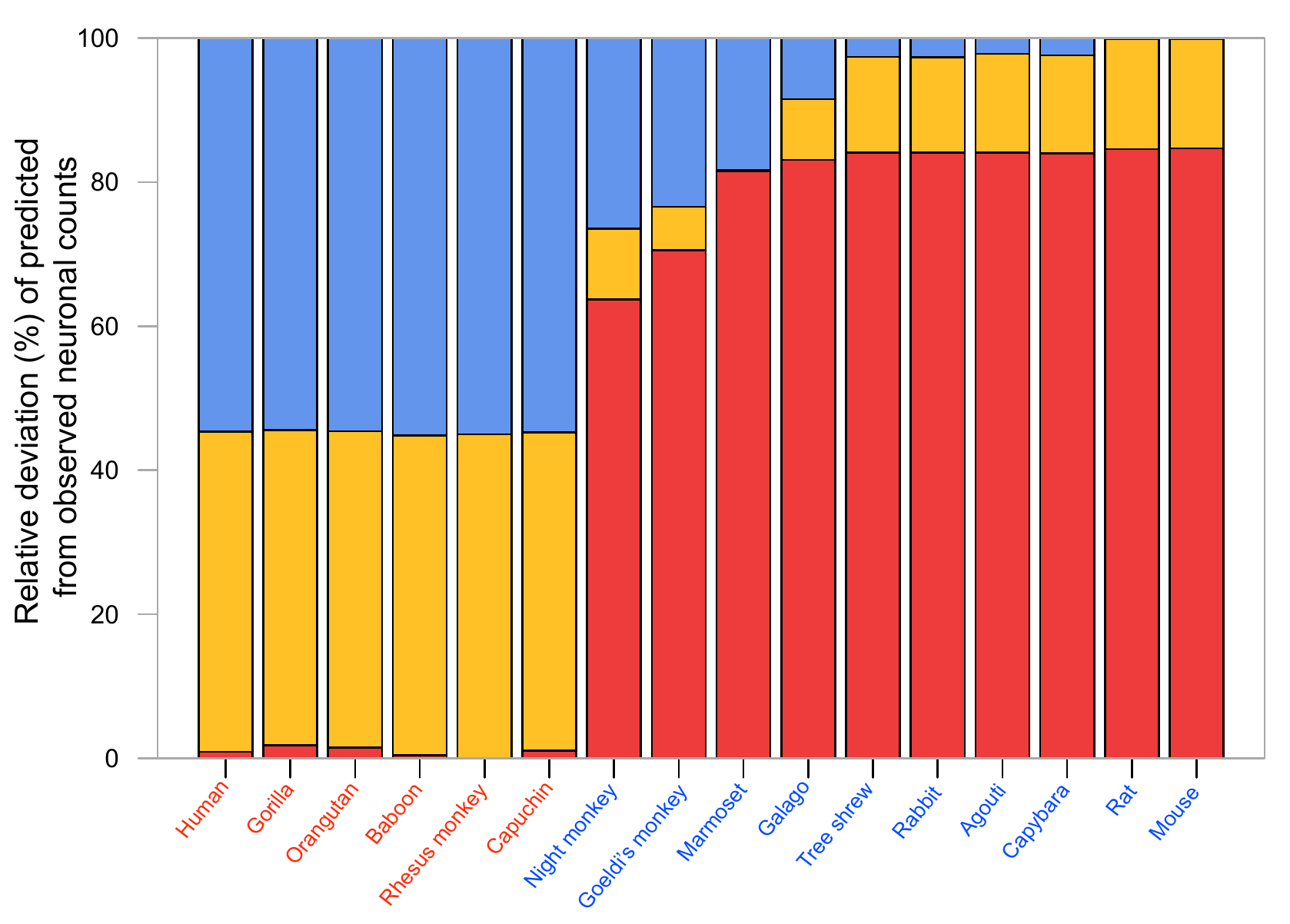}}
\caption*{\textbf{Figure S7:} Stacked barplot, for the indicated species, of deviations between the observed neocortical neuron counts and the ones predicted based on human (red), mouse (blue) and marmoset (yellow) lineage combinations (see Table 2 and Figure 5). For each species, deviations were calculated as $|$100*((Predicted –- Observed)/Observed)$|$ and then divided by the sum of deviations obtained for all three lineage combinations. Note that predictions based on the marmoset lineage combination deviate from observed neuron counts not only for the 6 species with a GI value $>$ 1.5 (red text), but also for 8 of the 10 species with a GI $\leq$ 1.5 (blue text), indicating a necessity for differential proportional occurrences of bRG in low-GI species. It is worth noting that natural intraspecific variation in neocortical neuron number has been shown to be considerably less than interspecific variation \cite{collins_neuron_2010,young_cell_2013}.}
\label{Figure S7}
\end{figure} 

\begin{figure}[hb]
\centering
\makebox[\textwidth][c]{
\includegraphics[scale=0.60]{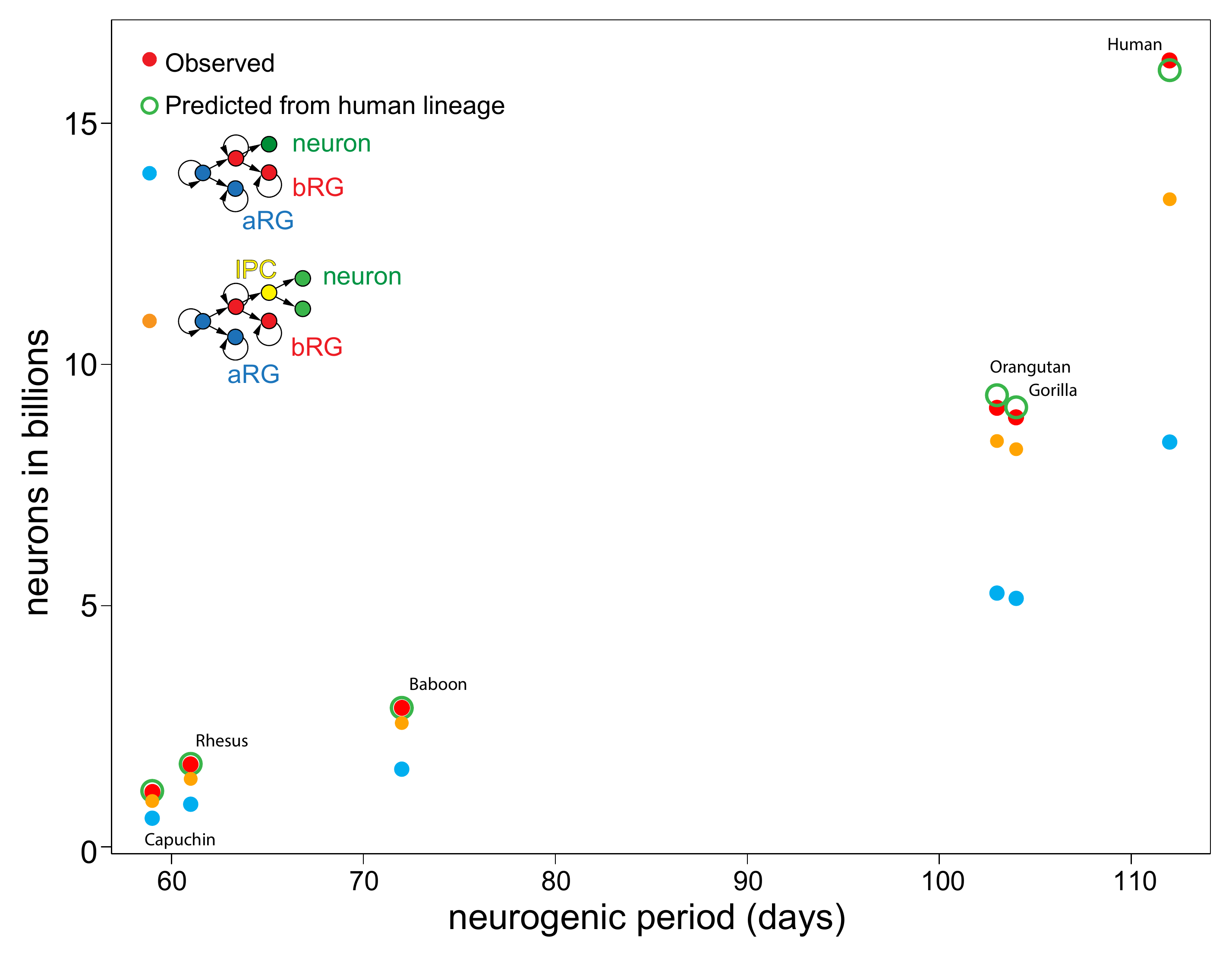}}
\caption*{\textbf{Figure S8:} Plot of observed neocortical neuronal count (red circles) as a function of neurogenic period for six species with a GI value $>$ 1.5. Predicted neuron counts are presented for the human lineage combination (green circles; see Figure 5, Table 2) and for two further lineages, each of which is assumed to have a 100$\%$ proportional occurrence: direct neurogenesis from bRG (blue circle) and indirect neurogenesis from bRG via a self-consuming IP cell (orange circle). Note that indirect neurogenesis from bRG via IPs is nearly sufficient to achieve the observed neuronal count in the Capuchin monkey.}
\label{Figure S8}
\end{figure} 

\begin{figure}[hb]
\centering
\makebox[\textwidth][c]{
\includegraphics[scale=0.65]{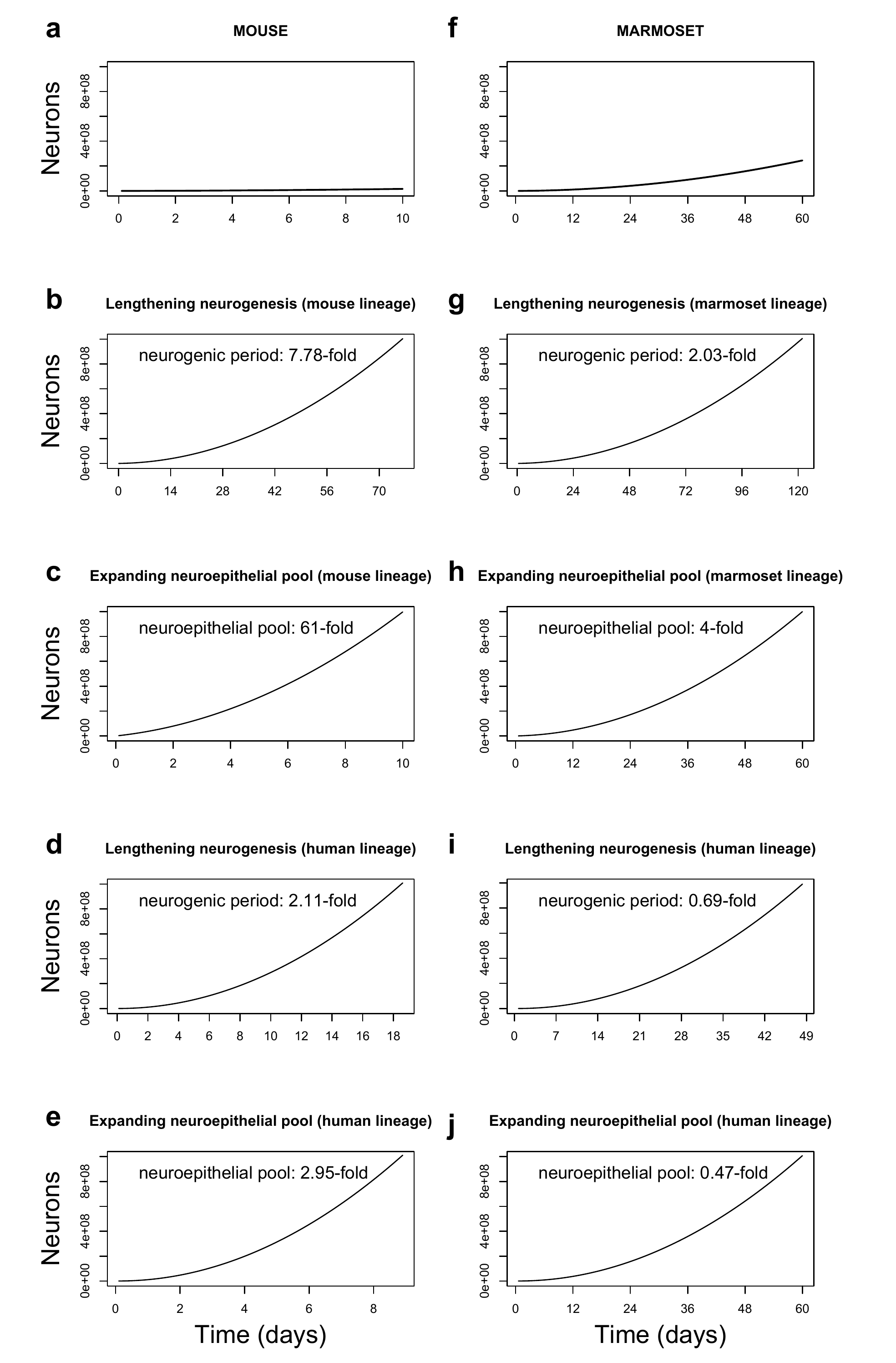}}
\caption*{\textbf{Figure S9:} Calculating the adaptiveness of proliferative basal progenitors in mouse (a-e) and marmoset (f-j) in achieving 10$^9$ neurons with respect to lengthening neurogenic period and expanding neuroepithelial founder pool size. The fold-change of lengthening neurogenic period or expanding neuroepithelial founder pool size is indicated in each relevant plot.
(a) The observed neurogenic period and founder pool size in mouse generates 1.37 x 10$^7$ neurons using the mouse lineage combination. 
(b, c) Lengthening the neurogenic period (b) or expanding the founder pool size (c) using the mouse lineage combination to achieve 10$^9$ neurons. 
(d, e) Lengthening the neurogenic period (d) or expanding the founder pool size (e) using the human lineage combination to achieve 10$^9$ neurons. 
(f) The observed neurogenic period and founder pool size in marmoset generates 2.45 x 10$^8$ neurons using the marmoset lineage combination. 
(g, h) Lengthening the neurogenic period (g) or expanding the founder pool size (h) using the marmoset lineage combination to achieve 10$^9$ neurons.
(i, j) Lengthening the neurogenic period (i) or expanding the founder pool size (j) using the human lineage combination to achieve 10$^9$ neurons.}
\label{Figure S9}
\end{figure} 

\begin{figure}[!h]
\centering
\makebox[\textwidth][c]{
\includegraphics[scale=1]{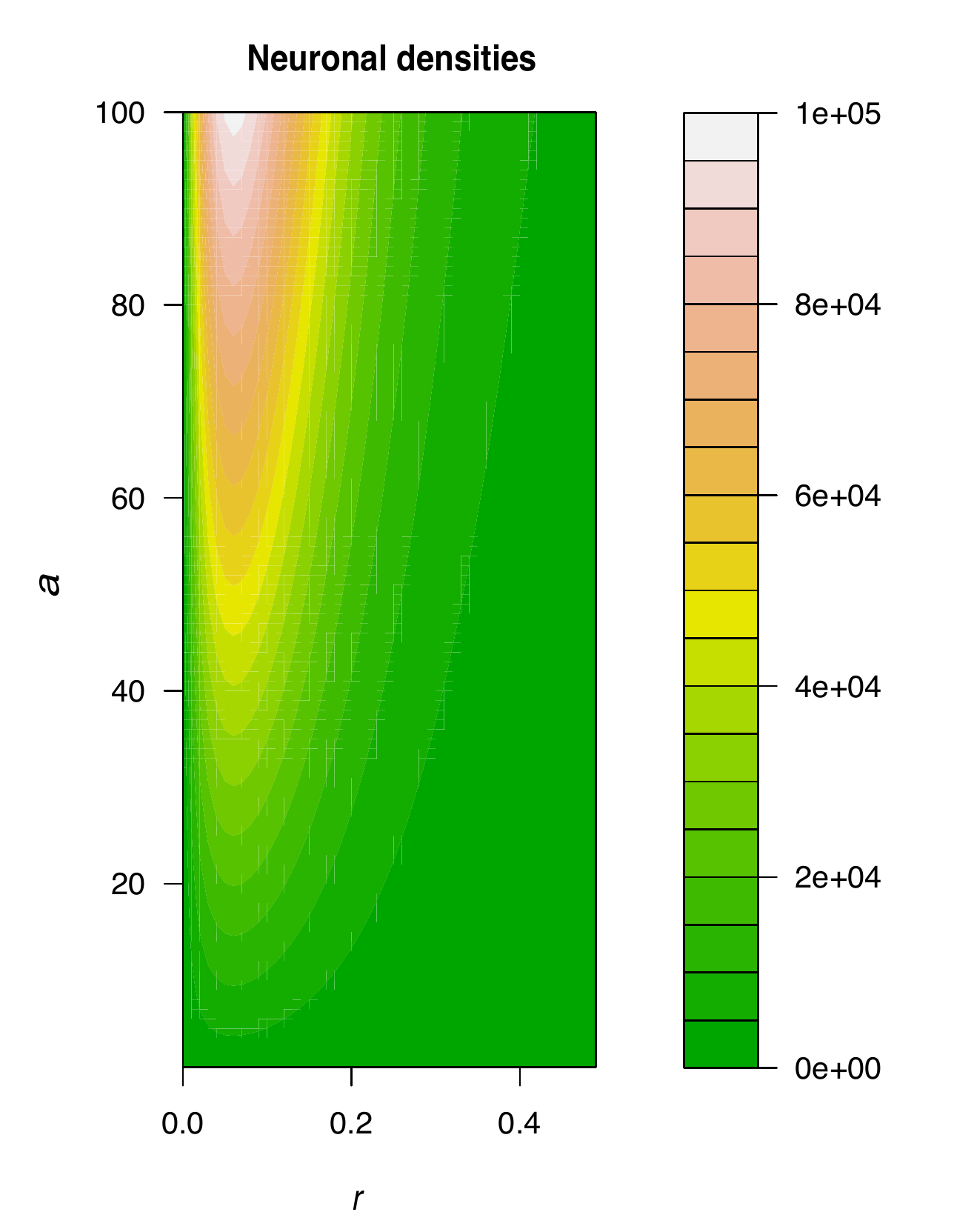}}
\caption*{\textbf{Figure S10:} Neuronal outputs from solutions to ODEs describing direct versus indirect neurogenesis for growth-rate values $\leq$ 0.5. Contour plot of neuronal densities for a varying initial asymmetrically dividing cell population (\textit{a}) and likelihood of direct (\textit{r} = 1) versus indirect (\textit{r} = 0) neurogenesis. Note that neuronal output increases maximally when both the initial cell pool increases (\textit{a} $\rightarrow$ 100) and the likelihood of indirect neurogenesis increases (\textit{r} $\rightarrow$ 0).}
\label{Figure S10}
\end{figure} 

\begin{figure}[hb]
\centering
\makebox[\textwidth][c]{
\includegraphics[scale=0.65]{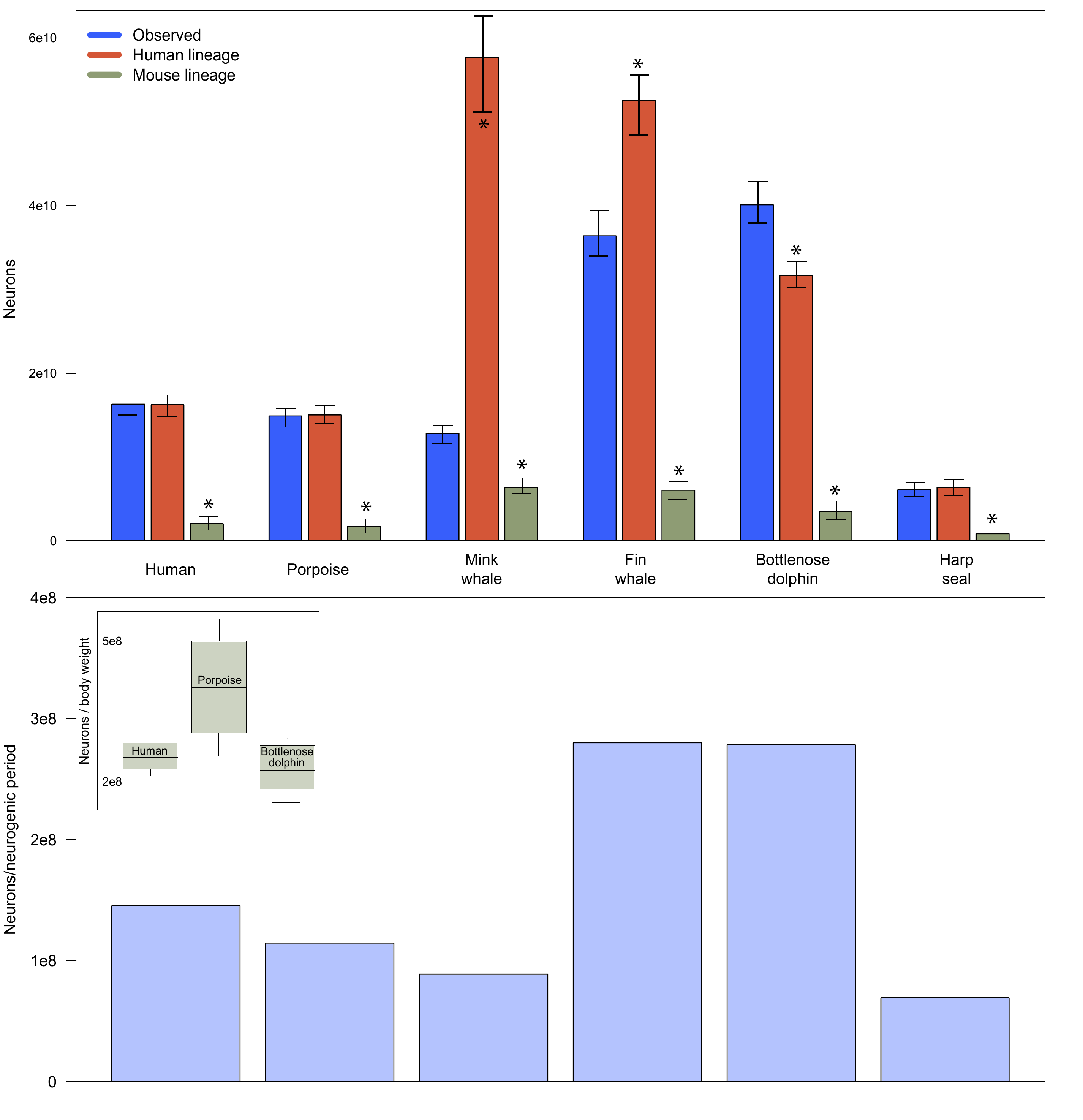}}
\caption*{\textbf{Figure S11:} Neocortical development in marine mammals may be largely explained by the same neurogenic program as terrestrial mammals. (a) Observed neocortical neuron numbers for human, four cetacean species, and one marine carnivore are shown beside neuron numbers calculated from the human (red) and mouse (green) lineages (see Text). Asterisks denote neuron numbers that are significantly different (T $>$ 7, P $<$ 0.05) from the observed. (b) The number of neurons generated per neurogenic day in the six species in (a). Inset: The total number of neocortical neurons as a proportion of body weight in human and two cetacean species. Note that the Bottlenose dolphin is the only species for which the human lineage is not sufficient to achieve its observed number of neurons. See Table S7 for data.} 
\label{Figure S11}
\end{figure} 

\begin{figure}[hb]
\centering
\makebox[\textwidth][c]{
\includegraphics[scale=0.50]{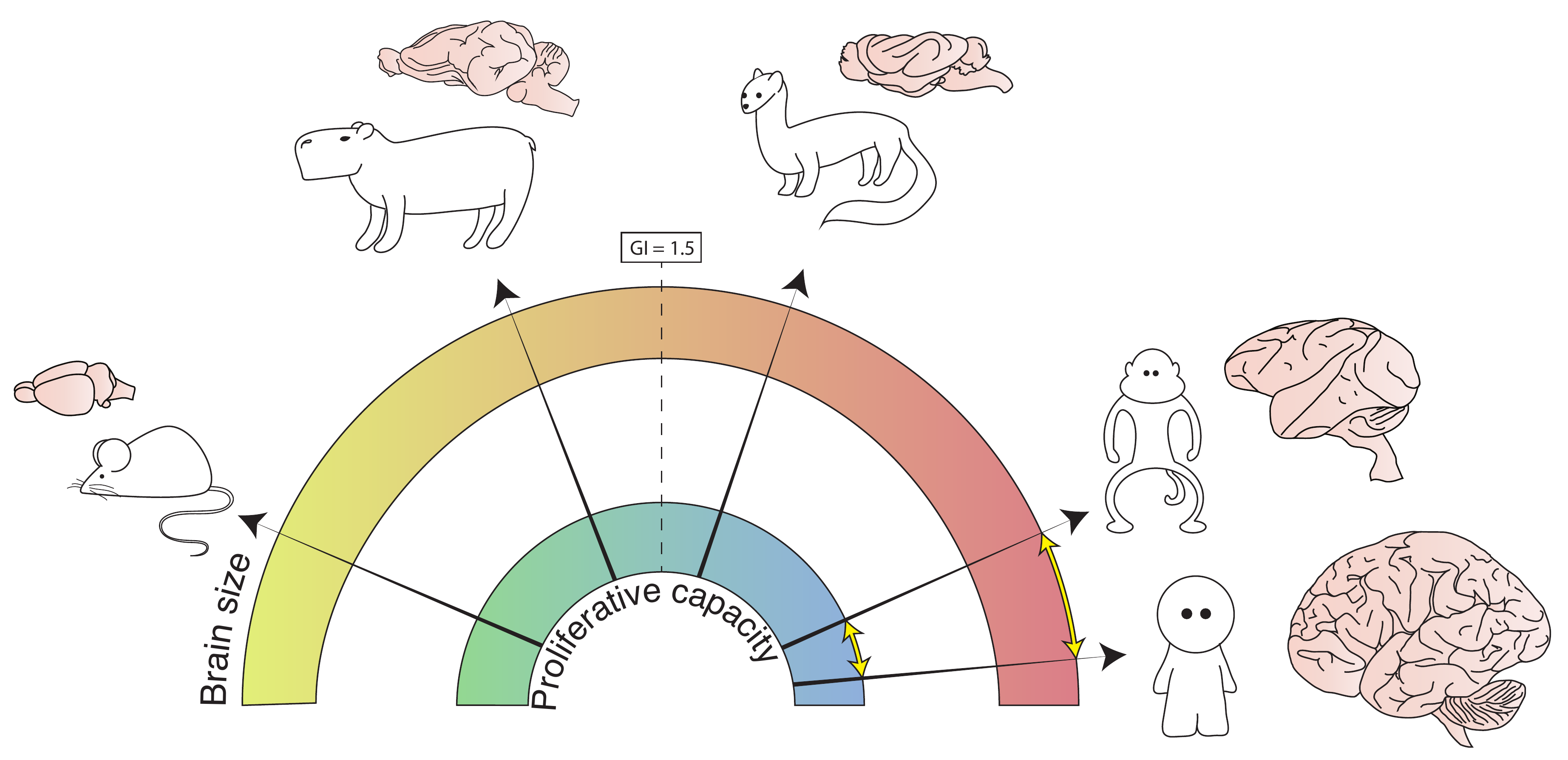}}
\caption*{\textbf{Figure S12:} Neocortical complexity, represented here as cortical gyrification, is tightly linked to progenitor behavior in the OSVZ. The nature of the link, however, is such that incremental changes to OSVZ progenitor behavior (inner ring) may effect exponential changes in neocortical complexity (outer ring). Therefore, minor changes in the proliferative capacity of basal progenitors (yellow arrow, inner ring) is needed to distinguish the major differences in neocortical complexity (yellow arrow, outer ring) between the macaque and human. It remains to be shown whether shifts in the proliferative capacity of OSVZ progenitors and neocortical complexity can occur independently (i.e., whether the arrow can be bent). Pictured clockwise: mouse, capybara, ferret, macaque, human.}
\label{Figure S11}
\end{figure} 

\begin{figure}[hb]
\centering
\makebox[\textwidth][c]{
\includegraphics[scale=0.55]{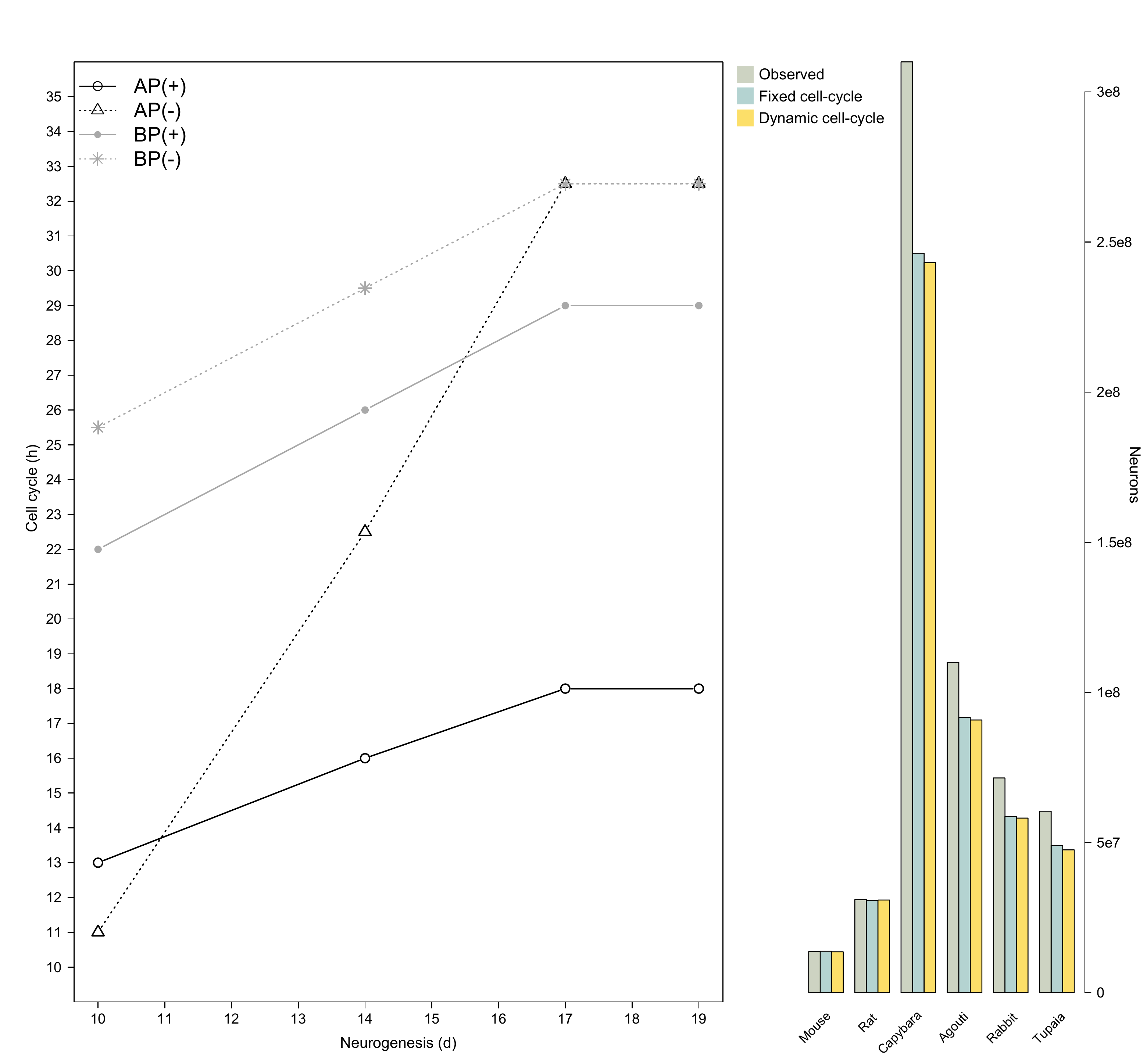}}
\caption*{\textbf{Figure S13:} Cell-cycle dynamics of progenitors in non-primates. (a) Cell-cycle for $Tis21\pm$ apical and basal progenitors at different stages of neurogenesis from live-imaging studies performed in the mouse \cite{arai_neural_2011}. (b) Barplot of the observed number of neurons in the neocortex of five rodents and a sister species to primates compared to the number of neurons predicted using a fixed cell-cycle of 18.5 hours, as was done in Figure 5, and the number of neurons predicted using dynamic cell-cycles for each progenitor as shown in (a). Note that for all species the predictions based on fixed and dynamic cell-cycles deviate by $< 1\%$. The percentage deviations between observed and mouse lineage-predicted neuron numbers are listed in Table S3.}
\label{Figure S13}
\end{figure}

\end{document}